\documentclass[11pt]{article}

\usepackage[utf8]{inputenc}
\usepackage[T1]{fontenc}
\usepackage[margin=1in]{geometry}
\usepackage{amsmath,amssymb}
\usepackage{graphicx}
\usepackage{booktabs}
\usepackage{caption}
\usepackage[round,authoryear]{natbib}
\usepackage[hidelinks]{hyperref}
\usepackage{multirow}
\usepackage{xcolor}
\usepackage[normalem]{ulem}

\graphicspath{{figs/}{../figs/}}
\title{\vspace{-1.5em}Continually learning neural-operator surrogate for\\
three-dimensional airborne electromagnetic Bayesian inversion}

\author{Jaehong Chung$^{1}$, Andrew Lockwood$^{2}$, and Jef Caers$^{1}$\\[3pt]
  \normalsize $^{1}$Mineral-X, Department of Earth and Planetary Sciences, Stanford University, USA\\[1pt]
  \normalsize $^{2}$Xcalibur Smart Mapping, Perth, Australia}
\date{}

\begin{document}
\maketitle

\begin{abstract}
\noindent
Three-dimensional probabilistic inversion of time-domain airborne electromagnetic (AEM) data is limited by the cost of the forward solve. Even though one simulation takes only tens of seconds, a Bayesian inversion of a survey of millions of soundings requires of order $10^{10}$ forward evaluations. To address this, we develop a continually learning neural-operator surrogate of the three-dimensional AEM forward operator that replaces the solver inside the Bayesian inversion. We start from the point of view that regardless of what geological prior is specified, Maxwell's laws remain invariant. Secondly, we avoid the limitation of learning on a single prior by continual learning on consecutive priors, which means our surrogate becomes richer as it is applied in future case studies, either by the authors, or by the scientific community. We use a validity check built on ensemble disagreement to divert cases with measurements outside the training range to the solver. Driven by the surrogate, the identical Markov chain Monte Carlo sampler reproduces the full-solver posterior, and its credible intervals cover the truth within 2.6 percentage points. Applied to the 2013 Capricorn TEMPEST survey in Western Australia, the surrogate inverts over two million soundings in seconds, a computation infeasible for the solver. Testing the geological prior against the entire survey costs minutes. The framework delivers uncertainty-quantified conductivity imaging at survey scale, which we believe is essential to perform near real-time mineral-systems targeting with geophysics.

\vspace{6pt}
\noindent\textbf{Highlights}
\begin{itemize}\setlength\itemsep{1pt}
\item One neural-operator surrogate accumulates geological priors, so the airborne electromagnetic surveys it can invert grow with every prior learned, and accuracy on the earlier priors improves.
\item The surrogate reproduces the full-solver posterior with calibrated credible intervals.
\item The surrogate inverts the two million soundings of the Capricorn TEMPEST survey in seconds, against about 26{,}300 years for the solver.
\end{itemize}

\vspace{6pt}
\end{abstract}

\section{Introduction}
\label{sec:intro}
Airborne electromagnetic (AEM) surveys sense the electrical conductivity of the subsurface and are a primary tool for critical-mineral exploration, groundwater assessment, and structural mapping \citep{nabighian1979quasi, auken2015overview, yang2012three}. The standard product of such a survey is a single conductivity model obtained by deterministic inversion. Many conductivity models, however, fit the same data within noise, because the electromagnetic inverse problem is non-unique \citep{tarantola1982generalized, tarantola2005inverse}. Exploration decisions based on a single model do not address the cost risks associated with follow-up surveys or drilling.

A Bayesian framework quantifies this non-uniqueness by sampling. It returns a posterior distribution of conductivity models consistent with the data and the prior, and the spread of that posterior is the uncertainty \citep{mosegaard1995monte, sambridge2002monte}. Its computational cost in three dimensions has kept it from routine use, for two reasons. One is the dimensionality of a voxelized model space. This can be addressed by parameterizing the conductivity field as a Gaussian process, which encodes spatial continuity in a small set of hyperparameters \citep{kitanidis1996geostatistical, wang2022hierarchical}, or any other set of lower-dimensional projections. The other is the number of forward solves the sampler requires. A three-dimensional electromagnetic solve takes tens of seconds, a sampler requires thousands of solves per sounding, and an airborne survey may have hundreds of thousands to millions of soundings. Replacing the solver with a forward surrogate reduces this cost by orders of magnitude while keeping an explicit likelihood, so accepted samples remain conditioned on the observations \citep{moghadas2020soil, yang2024efficient}. A network can instead be trained to carry the data to the model directly, which removes the sampler as well as the solver \citep{wu2023fast}. The posterior it returns is subject to the specific prior it is trained on. We take the other route and surrogate only the forward map, so the sampler and the likelihood are unchanged. A forward surrogate has recently reached three-dimensional ground-loop data \citep{liu2026probability} and survey-long sequences of two-dimensional lines \citep{caers2026stochastic}.

A limitation common to these forward surrogates is that the surrogate itself is static and prior-specific. It is trained once, on models drawn from a single geostatistical prior, and it is accurate only within that prior's support. But surveys do not need to remain within one prior. Geology changes along the flight lines, and repeated surveys extend the data in time as well as in space. When a sounding falls outside the training support, the surrogate still returns a prediction. The failure in extending the single prior is most severe for compact, strong conductors of economic interest, which are the structures a smooth training ensemble contains least. Training a separate surrogate for each new region does not settle the problem. Each new surrogate requires its own solver-labeled data, and a growing collection of single-prior models is not the same as a single surrogate that accumulates multiple priors in a single neural model. This distribution-shift problem is identified as an open direction in \citet{liu2026probability}.

We address the shift by learning the forward operator continually. The physics governing the electromagnetic response, Maxwell's equations, remains invariant regardless of the prior. In machine learning terms a new prior is a covariate shift. Continual learning methods built for conflicting tasks protect the network's weights from change \citep{kirkpatrick2017overcoming}. Between priors nothing conflicts, so it is enough to retain a small memory of models from the earlier priors and to require each update to reproduce the previous predictions on them, by experience replay and function-space distillation \citep{hinton2015distilling, li2017learning, lopez2017gradient}. One operator then accumulates priors rather than exchanging one for another, and an accuracy test accepts an update only if no earlier prior regresses.

The key contributions of this study are the following: (1) a neural operator that learns the three-dimensional airborne electromagnetic forward map across a sequence of geostatistical priors without forgetting, (2) a validity check that flags soundings outside the training support and routes them to the full solver, and (3) a demonstration that the surrogate reproduces the solver posterior on real data from Western Australia. The operator then inverts a full airborne survey, a computation infeasible for the traditional solver.

\section{AEM survey and geological priors}
\label{sec:data}

The inversion targets an airborne electromagnetic survey in Western Australia. One geological prior is built for the survey area, and four more are built for regions with different conductivity structures. The idea is to test the continual learning by a sequential assimilation of these various priors.

\subsection{The Capricorn survey, Western Australia}
\label{subsec:capricorn}

\begin{figure}[htbp]\centering
\includegraphics[width=0.95\linewidth]{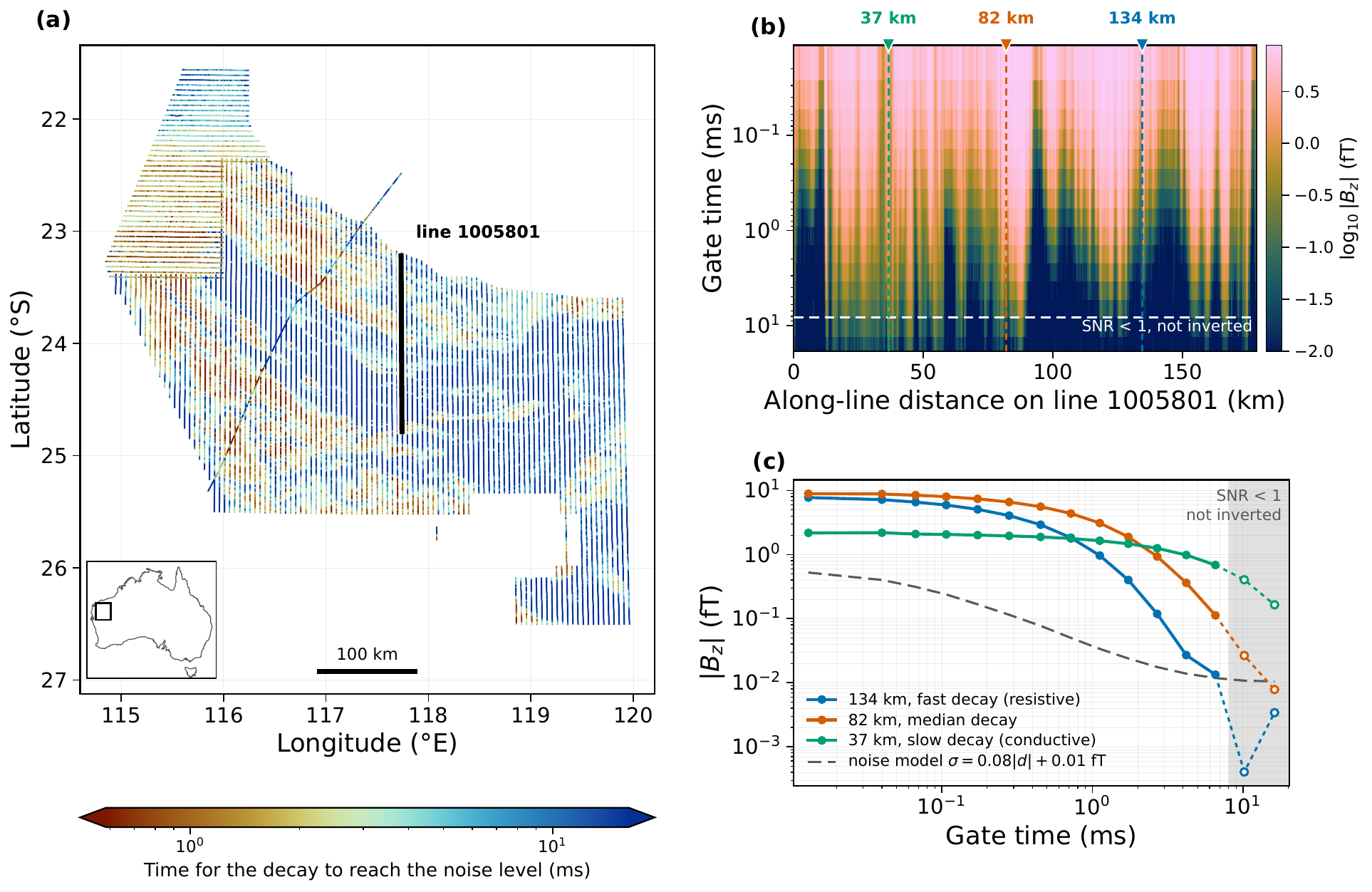}
    \caption{The Capricorn TEMPEST survey. (a) The 191 flight lines, colored by the gate time at which each sounding's decay reaches the noise level. Line 1005801 is marked in black, and the inset locates the survey in Australia. (b) Decay section of line 1005801, $\log_{10}|B_z|$ (fT) over the fifteen gates along the 178\,km line. Dashed vertical lines mark the three soundings of panel (c). (c) Decays of those three soundings, spanning slow (conductive), median, and fast (resistive) responses. The dashed gray curve is the noise model, and gates below it are shaded and drawn open.}
\label{fig:line}
\end{figure}

The field data are from the 2013 TEMPEST survey over the Turee Creek area, Western Australia, released as the Capricorn regional survey. It holds 2{,}155{,}272 soundings on 191 flight lines at 5\,km line spacing (Figure~\ref{fig:line}(a)). The fixed-wing system transmits at 25\,Hz and records the vertical magnetic field in fifteen time gates from 0.013 to 16.2\,ms. The nominal transmitter clearance is 120\,m, and the receiver trails 117\,m behind and 41.5\,m below the transmitter. Those three numbers define the survey geometry the surrogate reads at inference. The released channel is corrected to the nominal clearance rather than to the recorded per-sounding value. 

The processing report provides the survey's additive noise level per gate, from 0.0267\,fT at the first gate to 0.0070\,fT at gate 14, with a median of 0.0117\,fT over the fifteen gates \citep{cgg2014capricorn}. Line 1005801 carries the single-line results below and grounds the prior built for this area (Figure~\ref{fig:line}(b,c)).

The geological prior for this area is built from the regional geology. The stratigraphy is a three-layer sequence. Lateritic regolith overlies Turee Creek Group siliciclastics, which overlie folded, magnetite-bearing banded iron formation of the Hamersley Group, 2.6 to 2.4\,Ga, deformed by the Ophthalmia Orogeny \citep{thorne1996geology,krapez1996sequence}. The regolith is deeply weathered and therefore conductive \citep{anandpaine2002}, while fresh banded iron formation is resistive \citep{dentith2014geophysics}. That contrast sets the vertical conductivity structure the prior encodes. Figure~\ref{fig:priorgeol}(a) shows the sequence as a conceptual section, and Table~\ref{tab:regions} gives the hyperparameter ranges the prior is sampled over.

\subsection{Comparison regions and their geological priors}
\label{subsec:regions}

\begin{figure}[htbp]\centering
\includegraphics[width=\linewidth]{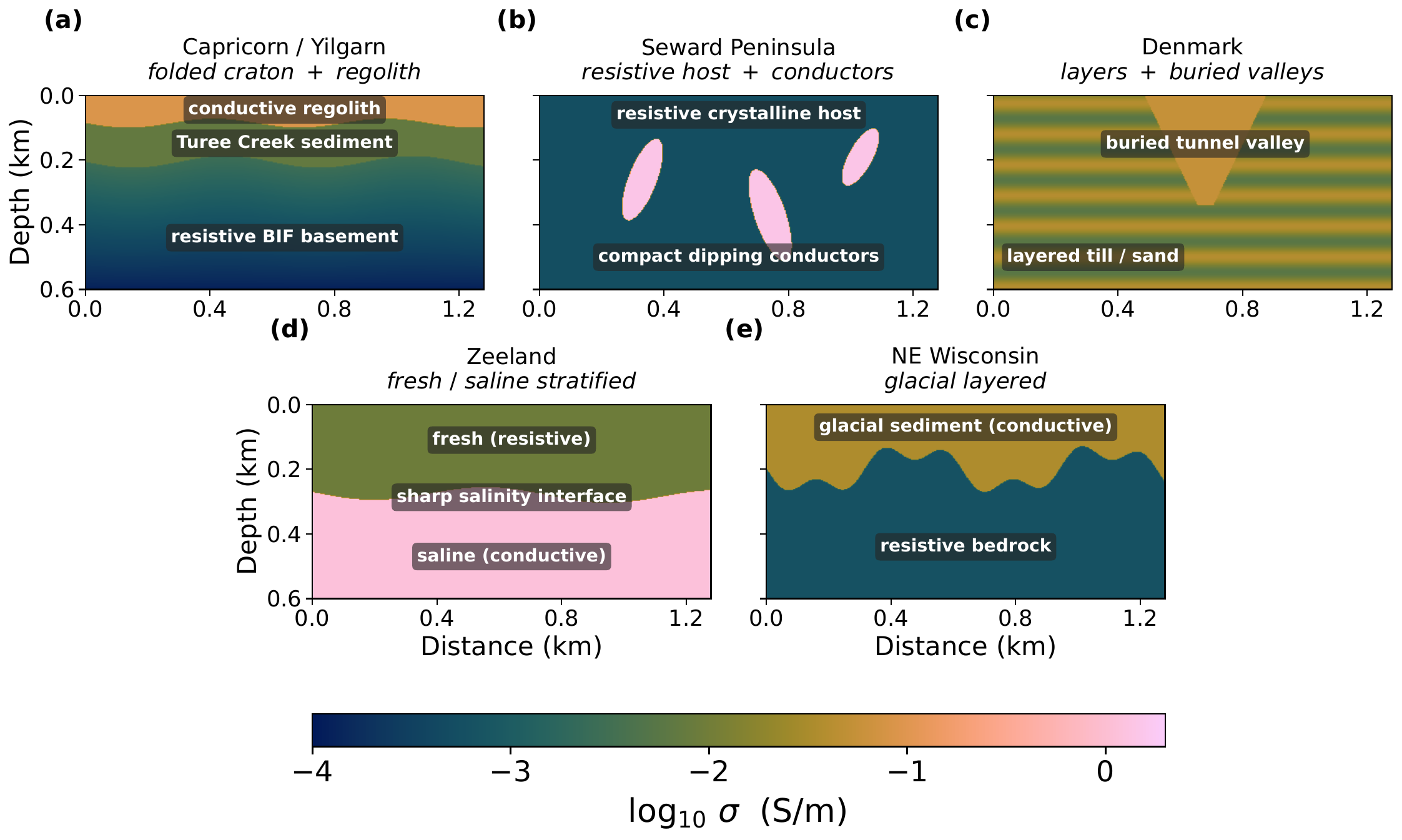}
    \caption{Conceptual geology of the five regions, on one shared $\log_{10}\sigma$ scale. (a) Capricorn, (b) Seward Peninsula, (c) Denmark, (d) Zeeland, and (e) northeastern Wisconsin.}
\label{fig:priorgeol}
\end{figure}

\begin{figure}[htbp]\centering
\includegraphics[width=0.96\linewidth]{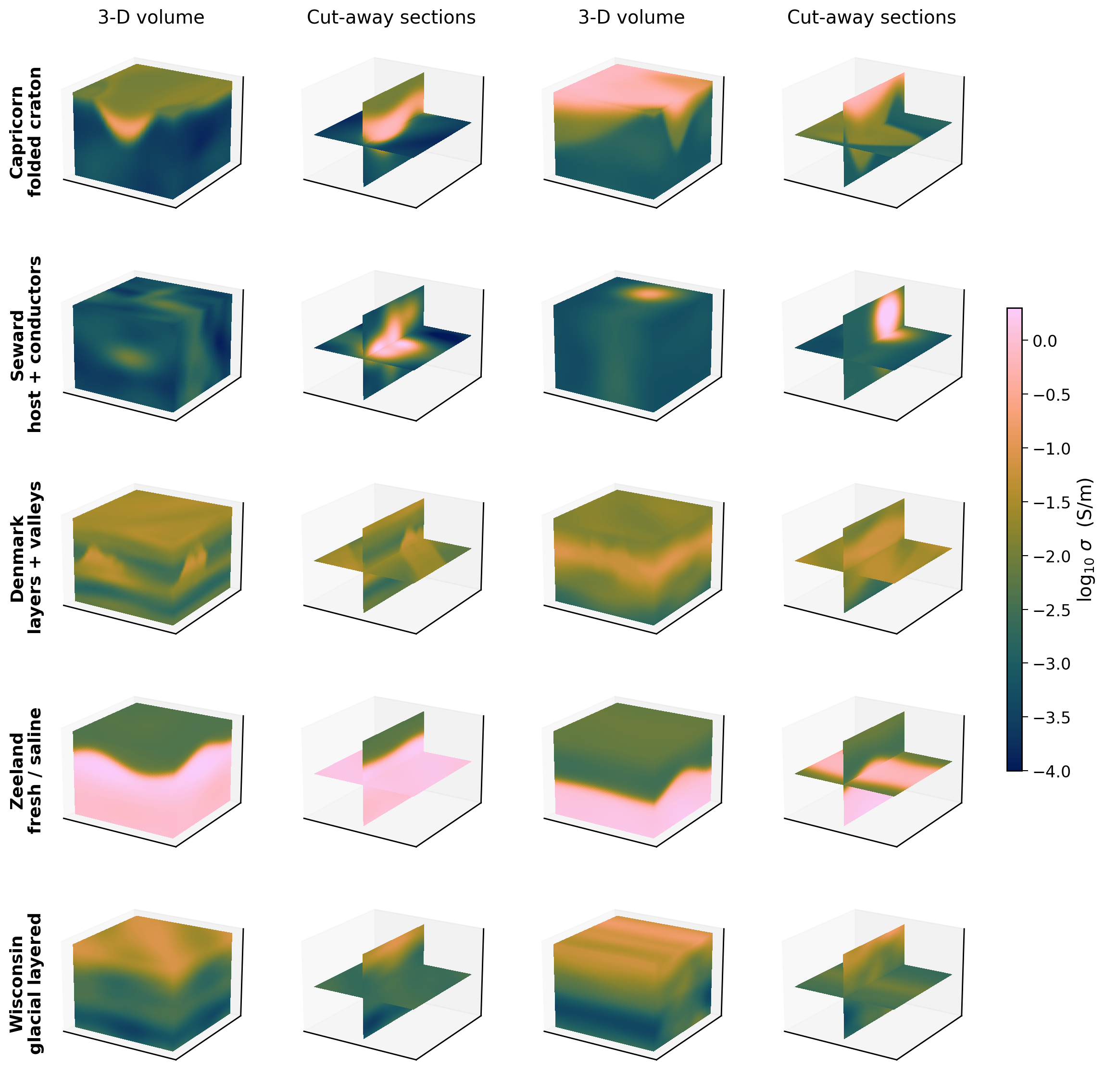}
    \caption{Two realizations of each geological prior, sampled from a Gaussian process. Each row is one prior, and each realization is shown as the three-dimensional volume and a cut-away along a horizontal and a vertical mid-plane.}
\label{fig:priordiv}
\end{figure}

One region comes with one geological prior, so the Capricorn survey alone cannot test whether one operator works across geology. We therefore build four more geological priors. The Seward Peninsula, Alaska, is a resistive crystalline host carrying compact graphite and sulfide conductors \citep{Emond2024Seward}. Denmark holds layered glacial sediments incised by buried tunnel valleys \citep{barfod2016compiling,moller2009integrated}. Zeeland, the Netherlands, is a coastal section with a sharp fresh over saline interface \citep{revil2017complex}. Northeastern Wisconsin holds conductive glacial cover over resistive bedrock \citep{minsley2022airborne}. Figure~\ref{fig:priorgeol}(b-e) shows the conceptual geology of each, beside the Capricorn section on the same scale. The priors are built from this geological information alone, and no survey data from these regions are used. Table~\ref{tab:regions} lists the hyperparameter ranges the four priors are sampled over. Figure~\ref{fig:priordiv} shows two realizations of each prior.

\begin{table}[htbp]\centering
\caption{Gaussian process hyperparameters of the five geological priors. Each entry is a uniform distribution $U[a,b]$ sampled per realization. $\sigma$ is the unit conductivity in $\log_{10}$ S/m, $h$ is a layer thickness, $d$ is a top or interface depth, and $w$ is a body width, all in meters. $\ell_h$ and $\ell_v$ are the horizontal and vertical correlation lengths in meters, $\alpha$ is the horizontal-to-vertical anisotropy factor, and $\nu$ is the Mat\'ern smoothness, where a larger $\nu$ gives a smoother field. (The Zeeland source is a frequency-domain survey, and only its conductivity structure is used.)}
\label{tab:regions}
\footnotesize
\setlength{\tabcolsep}{3.5pt}
\renewcommand{\arraystretch}{1.15}
\begin{tabular}{@{}l l l l c c c c l@{}}
\toprule
Prior & Type & Variable & $U[a,b]$ & $\ell_h$ (m) & $\ell_v$ (m) &
$\alpha$ & $\nu$ & Source \\
\midrule
\multirow{5}{*}{Capricorn}
  & \multirow{3}{*}{Conductivity}
    & $\sigma_{\mathrm{regolith}}$ & $[-2.5,\,0]$    & \multirow{5}{*}{[300,\,900]}
    & \multirow{5}{*}{[120,\,300]} & \multirow{5}{*}{[1.0,\,2.5]}
    & \multirow{5}{*}{[2,\,5]}
    & \multirow{5}{*}{\shortstack[l]{\citet{thorne1996geology}\\ \citet{anandpaine2002}}} \\
  & & $\sigma_{\mathrm{sediment}}$ & $[-3.0,\,-0.1]$ & & & & & \\
  & & $\sigma_{\mathrm{basement}}$ & $[-4.0,\,-1.5]$ & & & & & \\
\cmidrule(lr){2-4}
  & \multirow{2}{*}{Geometry}
    & $h_{\mathrm{regolith}}$      & $[0,\,120]$     & & & & & \\
  & & $d_{\mathrm{basement}}$      & $[40,\,400]$    & & & & & \\
\midrule
\multirow{4}{*}{Seward}
  & \multirow{2}{*}{Conductivity}
    & $\sigma_{\mathrm{host}}$     & $[-3.3,\,-2.6]$ & \multirow{4}{*}{[300,\,900]}
    & \multirow{4}{*}{[120,\,300]} & \multirow{4}{*}{[1.0,\,3.0]}
    & \multirow{4}{*}{[1.5,\,4]}
    & \multirow{4}{*}{\citet{Emond2024Seward}} \\
  & & $\sigma_{\mathrm{body}}$     & $[-0.5,\,0.5]$  & & & & & \\
\cmidrule(lr){2-4}
  & \multirow{2}{*}{Geometry}
    & $w_{\mathrm{body}}$          & $[160,\,480]$   & & & & & \\
  & & $d_{\mathrm{body}}$          & $[80,\,340]$    & & & & & \\
\midrule
\multirow{4}{*}{Denmark}
  & \multirow{3}{*}{Conductivity}
    & $\sigma_{\mathrm{top}}$      & $[-1.7,\,-1.3]$ & \multirow{4}{*}{[400,\,1200]}
    & \multirow{4}{*}{[70,\,150]}  & \multirow{4}{*}{[2,\,5]}
    & \multirow{4}{*}{[1.5,\,4]}
    & \multirow{4}{*}{\shortstack[l]{\citet{barfod2016compiling}\\ \citet{moller2009integrated}}} \\
  & & $\sigma_{\mathrm{base}}$     & $[-2.6,\,-2.0]$ & & & & & \\
  & & $\sigma_{\mathrm{valley}}$   & $[-1.5,\,-1.1]$ & & & & & \\
\cmidrule(lr){2-4}
  & Geometry
    & $d_{\mathrm{valley}}$        & $[90,\,240]$    & & & & & \\
\midrule
\multirow{3}{*}{Zeeland}
  & \multirow{2}{*}{Conductivity}
    & $\sigma_{\mathrm{fresh}}$    & $[-2.3,\,-1.5]$ & \multirow{3}{*}{[700,\,2500]}
    & \multirow{3}{*}{[150,\,320]} & \multirow{3}{*}{[1.0,\,2.0]}
    & \multirow{3}{*}{[2,\,5]}
    & \multirow{3}{*}{\citet{revil2017complex}} \\
  & & $\sigma_{\mathrm{saline}}$   & $[-0.6,\,0.1]$  & & & & & \\
\cmidrule(lr){2-4}
  & Geometry
    & $d_{\mathrm{interface}}$     & $[90,\,340]$    & & & & & \\
\midrule
\multirow{3}{*}{Wisconsin}
  & \multirow{2}{*}{Conductivity}
    & $\sigma_{\mathrm{cover}}$    & $[-1.6,\,-1.1]$ & \multirow{3}{*}{[400,\,1500]}
    & \multirow{3}{*}{[90,\,190]}  & \multirow{3}{*}{[1.5,\,3.5]}
    & \multirow{3}{*}{[1.5,\,4]}
    & \multirow{3}{*}{\citet{minsley2022airborne}} \\
  & & $\sigma_{\mathrm{bedrock}}$  & $[-3.0,\,-2.3]$ & & & & & \\
\cmidrule(lr){2-4}
  & Geometry
    & $d_{\mathrm{bedrock}}$       & $[150,\,360]$   & & & & & \\
\bottomrule
\end{tabular}
\end{table}

\section{Methodology}
\label{sec:method}
We introduce the methodology to build a continually learning surrogate of the three-dimensional forward solver and the Bayesian inversion that uses it. The surrogate learns the map from a conductivity model to its airborne decay, $\sigma\mapsto\mathbf{d}$, over a sequence of geological priors. At inference a validity check decides whether a sounding lies inside the range the surrogate was trained on and passes it to the full solver when it does not. The surrogate then replaces the solver inside a Markov chain Monte Carlo sampler based on the probability perturbation method (PPM), which recovers the posterior distribution of conductivity models for each sounding. We use PPM with a Metropolis sampler because it is agnostic to the forward model, so the identical sampler runs with the solver or the surrogate. Figure~\ref{fig:method} summarizes the framework and each subsection provides the details.

\subsection{Governing equations for forward simulation}
\label{subsec:governing}
Airborne measurements record a diffusive process. When the transmitter current is switched off, the collapsing primary field induces eddy currents in the subsurface, and the current system diffuses downward and outward while it decays \citep{nabighian1979quasi}. The rate of decay provides the subsurface conductivity information. A conductive subsurface sustains the current system and the measured field decays slowly, whereas a resistive subsurface dissipates it within the earliest gates. The receiver records the vertical magnetic field of this decay at fifteen gate times.

The governing equations are Maxwell's equations in the quasi-static limit. At the time scales and subsurface conductivities of airborne surveys, conduction currents dominate displacement currents, so Faraday's law and Amp\`ere's law reduce to
\begin{equation}
    \nabla \times \mathbf{E}(\mathbf{x}, t) = -\frac{\partial \mathbf{B}(\mathbf{x}, t)}{\partial t}, \qquad
    \nabla \times \mathbf{B}(\mathbf{x}, t) = \mu_0 \big( \sigma(\mathbf{x}) \, \mathbf{E}(\mathbf{x}, t) + \mathbf{J}_s \big),
    \label{eq:maxwell}
\end{equation}
where $\mathbf{E}$ and $\mathbf{B}$ are the electric and magnetic fields, $\sigma$ is the electrical conductivity, $\mu_0$ is the vacuum permeability, and $\mathbf{J}_s$ is the source current of the transmitter loop. Taking the curl of Faraday's law and substituting Amp\`ere's law eliminates $\mathbf{B}$ and gives the diffusion equation for the electric field,
\begin{equation}
    \nabla \times \nabla \times \mathbf{E} + \mu_0 \, \sigma(\mathbf{x}) \, \frac{\partial \mathbf{E}}{\partial t} = -\mu_0 \frac{\partial \mathbf{J}_s}{\partial t},
    \label{eq:diffusion}
\end{equation}
solved for a step-off source, with the pre-shutoff steady state as the initial condition and the fields vanishing at infinity. The magnetic field follows from Faraday's law,
\begin{equation}
    \mathbf{B} = -\int \nabla \times \mathbf{E} \, dt,
    \label{eq:faraday_b}
\end{equation}
and is sampled at the receiver location and gate times.

The instrument does not record a step-off response. The TEMPEST system transmits a 50\% duty-cycle square wave at 25~Hz, and the released data are transformed to the response of a 100\% duty-cycle waveform at the same base frequency. Because the governing equation is linear in the fields, the duty-cycle response is the alternating-sign superposition of step-off responses across successive polarity reversals,
\begin{equation}
    d(t) = \sum_{k=0}^{K} (-1)^k \, s\!\left(t + k \, T_{1/2}\right), \qquad T_{1/2} = 20 \text{ ms},
    \label{eq:dutycycle}
\end{equation}
where $d(t)$ is the recorded response, $s(\cdot)$ is the step-off response, $T_{1/2}$ is the half-period, and $K$ is the number of half-cycles retained. The sum converges within a few half-cycles.

The response of one sounding is laterally local. The induced current system contributes measurably to the data only within a bounded lateral zone, the footprint of the sounding, of order one kilometer for this system and depth range. Each sounding is therefore modeled on a local crop of the survey volume, $32 \times 32 \times 24$ voxels at $40 \times 40 \times 25$~m. The simulation is a three-dimensional finite-volume solve on an OcTree mesh in SimPEG \citep{cockett2015simpeg}. 

\subsection{Neural operator surrogate for forward simulation}
\label{subsec:surrogate}
The surrogate approximates the solution operator of Equations~\ref{eq:diffusion} and \ref{eq:faraday_b}, which maps a conductivity field to the decay it produces,
\begin{equation}
    F : \sigma(\mathbf{x}) \mapsto \mathbf{d} = \big( B_z(t_1), \ldots, B_z(t_{15}) \big).
    \label{eq:forwardmap}
\end{equation}
The operator is defined on functions rather than fixed vectors. The network inherits this structure, taking the conductivity as an input field and outputting the response at any gate time and geometry in the training range. The architecture below is therefore a neural operator rather than a regression network, trained on solver-labeled pairs $(\sigma, \mathbf{d})$. The network has three parts, each with a specific purpose. A model encoder compresses the conductivity field into coefficients. A prior embedding tells the network which geological prior the field was drawn from. A query network reads the gate time and the survey geometry and builds the basis the coefficients weight. We will see that this allows changing the survey geometry itself when applying the surrogate model. Figure~\ref{fig:method}(a,b) shows the three parts on one forward pass, with the tensors each stage actually produces, and the following paragraphs describe each component in detail.

First, the model encoder is a three-dimensional Fourier Neural Operator \citep{li2020fourier, kovachki2023neural}. The input is the conductivity field on the $32 \times 32 \times 24$ crop, concatenated with three normalized coordinate channels, and a pointwise linear layer lifts this four-channel field to a $C$-channel feature field $v_0$. Each of the $L$ layers then filters the field in the frequency domain, keeps the eight lowest modes per axis, and adds a pointwise map,
\begin{equation}
    v_{l+1} = \phi \big( W_l v_l + \mathcal{F}^{-1}( R \odot \mathcal{F} v_l ) \big), \qquad l = 0, \ldots, L-1,
    \label{eq:fno}
\end{equation}
where $v_l$ is the $C$-channel feature field at layer $l$, $\mathcal{F}$ is the Fourier transform over the three spatial axes, $R$ holds the learnable weights that mix the channels of each retained low mode, $\odot$ denotes the mode-wise multiplication, $W_l$ is a pointwise linear map, and $\phi$ is a GELU nonlinearity. Averaging the final field $v_L$ over the voxel grid $\Omega$ of the crop gives the encoder feature,
\begin{equation}
    h = \frac{1}{|\Omega|} \sum_{\mathbf{x} \in \Omega} v_L(\mathbf{x}) \; \in \mathbb{R}^{C},
    \label{eq:pooling}
\end{equation}
where $|\Omega| = 32 \times 32 \times 24$ is the number of voxels.

Second, the prior embedding is a compact code $z$, a fingerprint of the prior, computed from a set of example realizations by a permutation-invariant encoder (DeepSets). The code modulates the encoder feature by a rescaling and a shift (feature-wise linear modulation, FiLM), and a two-layer perceptron turns the modulated feature into $P$ coefficients,
\begin{equation}
    z = \rho \Big( \tfrac{1}{E} \sum_{e} \psi(\varphi_e) \Big), \qquad
    h' = \gamma(z) \odot h + \beta(z), \qquad
    \mathbf{b} = W_2 \, \phi \big( W_1 h' \big) \; \in \mathbb{R}^{P},
    \label{eq:embedding}
\end{equation}
where $\varphi_e$ is a six-number statistic computed from each of the $E$ example realizations, holding the mean, the standard deviation, the horizontal and vertical correlation lengths, the conductive fraction, and the vertical contrast of the prior, $\psi$ and $\rho$ are the encoder networks, $\gamma, \beta$ are the modulation scale and shift, and $W_1, W_2$ are the weights of the projection network. The coefficients $\mathbf{b} = (b_1, \ldots, b_P)$ depend on the conductivity field through $h$ and on the prior through $z$, written $b_p(\sigma; z)$ below. Because the governing physics is consistent regardless of the prior, the embedding is not meant to change the operator the network approximates. It lets a network of fixed size allocate its capacity to the part of model space each prior occupies.

Third, the query network is a small perceptron $\tau$ that reads the query $q = [\log_{10} t, \; \mathbf{g}]$ of gate time $t$ and the three-component survey geometry $\mathbf{g}$, the transmitter clearance and the in-line and vertical receiver offsets, and returns the basis,
\begin{equation}
    \boldsymbol{\tau}(q) = \big( \tau_1(q), \ldots, \tau_P(q) \big) \; \in \mathbb{R}^{P}.
    \label{eq:query}
\end{equation}
The transmitter clearance changes from sounding to sounding as the aircraft follows the terrain, and each sounding is inverted at its own recorded geometry rather than at one nominal height.

The three parts assemble into the prediction, the DeepONet form of \citet{lu2021learning}, with the model encoder as its branch and the query network as its trunk, optionally on top of an analytic layered-earth baseline,
\begin{equation}
    \hat{y}(q) = \log_{10} \big| \hat{d}(q) \big| = \log_{10} \big| d_{1\mathrm{D}}(\bar{\sigma}; q) \big| + \sum_{p=1}^{P} b_p(\sigma; z) \, \tau_p(q) + b_0,
    \label{eq:deeponet}
\end{equation}
where $d_{1\mathrm{D}}$ is a fast one-dimensional layered-earth forward of the footprint-averaged column $\bar{\sigma}(z)$, evaluated at the query $q$, and $b_0$ is a learned bias. The baseline carries positivity and the late-time behavior, and the branch-trunk sum supplies the three-dimensional correction. It is used in the constrained controlled-study checkpoint; the production and field checkpoints deployed here set it to zero and predict $\log_{10}|\hat{d}|$ directly. The coefficients $b_p$ are independent of the query, so they are computed once per model and reused across all gate times and geometries.

\subsection{Loss function with continual learning and physical constraints}
\label{subsec:continual}
The inversion aims to recover the posterior $p(\sigma \mid \mathbf{d}) \propto p(\mathbf{d} \mid \sigma) \, p(\sigma)$. Across a sequence of geostatistical priors, the likelihood $p(\mathbf{d} \mid \sigma)$ is fixed by the physics of Section~\ref{subsec:governing} and only the prior $p(\sigma)$ changes, so a new prior widens the conductivity fields the surrogate covers without changing the map it learns. The goal of training is one operator that stays accurate across geologically distinct priors.

First, the base loss term fits the surrogate to solver labels on the current prior $\pi_s$, where $s$ indexes the stage in the sequence,
\begin{equation}
    \mathcal{L}_{\text{data}}^{\pi_s}(\theta) = \frac{1}{N n_g} \sum_{i=1}^{N} \sum_{t=1}^{n_g} \big( \hat{y}_{i,t} - y_{i,t} \big)^2,
    \label{eq:dataloss}
\end{equation}
where $y_{i,t}$ is the solver label of training model $i$ at gate $t$, $\hat{y}_{i,t}$ is the prediction of Equation~\ref{eq:deeponet} for the same model and gate, $N$ is the number of training models, and $n_g = 15$ is the number of gates.

Second, when the survey reaches a new geological region, training on the new prior alone overwrites what the surrogate learned on the earlier ones. Experience replay prevents the overwrite, a strategy inspired by the brain's replay of past experience during learning \citep{rolnick2019experience}. A memory $\mathcal{R}$ keeps a fixed number $M$ of models per earlier prior, drawn without replacement, and the replay term evaluates the data loss of Equation~\ref{eq:dataloss} on the memory,
\begin{equation}
    \mathcal{L}_{\text{replay}}(\theta) = \mathcal{L}_{\text{data}}^{\mathcal{R}}(\theta).
    \label{eq:replay}
\end{equation}
The memory grows only with the number of priors seen, by $M$ models per prior. A second term on the same memory requires the update to reproduce the previous checkpoint's own predictions there, not only the solver labels, which is function-space distillation,
\begin{equation}
    \mathcal{L}_{\text{distill}}(\theta, \theta_{s-1}) = \frac{1}{|\mathcal{R}| \, n_g} \sum_{i \in \mathcal{R}} \sum_{t=1}^{n_g} \big( \hat{y}_{i,t}(\theta) - \hat{y}_{i,t}(\theta_{s-1}) \big)^2 ,
    \label{eq:distill}
\end{equation}
where $\theta_{s-1}$ is the checkpoint the stage started from, held fixed.

Third, two physical consistency terms encode domain knowledge in the loss. A diffusive step-off field decays monotonically and its decay curve is concave on a log-log plot. The first term $\mathcal{L}_{\text{mono}}$ penalizes a predicted decay that rises with time, and the second term $\mathcal{L}_{\text{cvx}}$ penalizes a predicted decay that curves upward in log-log,
\begin{equation}
    \mathcal{L}_{\text{mono}} = \overline{\big( \Delta_t \hat{y} \big)_+^{2}}, \qquad
    \mathcal{L}_{\text{cvx}} = \overline{\big( \Delta_u^2 \hat{y} \big)_+^{2}},
    \label{eq:consistency}
\end{equation}
where $\Delta_t \hat{y}$ is the first difference of the predicted decay over gate time, $\Delta_u^2 \hat{y}$ is its second difference over $u = \log_{10} t$, $(x)_+ = \max(0, x)$ is the positive part, and the overbar is the mean over gates and models. The penalty applies to the predicted decay of the training data the network is trained on. That training data is the duty-cycle one of Equation~\ref{eq:dutycycle}, whose labels need not be monotone, so the two terms act as a soft bias rather than as a property the labels already satisfy.

Thus, the loss at stage $s$ assembles the terms,
\begin{equation}
    \mathcal{L}_s = \mathcal{L}_{\text{data}}^{\pi_s}(\theta) + \lambda_r \, \mathcal{L}_{\text{replay}}(\theta) + \lambda_d \, \mathcal{L}_{\text{distill}}(\theta, \theta_{s-1}) + \lambda_c \big( \mathcal{L}_{\text{mono}} + \kappa \, \mathcal{L}_{\text{cvx}} \big),
    \label{eq:totalloss}
\end{equation}
where $\lambda_r$ and $\lambda_d$ weight the two rehearsal terms and $\lambda_c$ the physical constraints. The rehearsal terms carry unit weight, $\lambda_r = \lambda_d = 1$, with one replay minibatch drawn per earlier prior at every step; the constraint terms are used in the single-stage controlled training, at $\lambda_c = 0.1$ and $\kappa = 0.3$, and no one run combines all four terms. Each stage starts from the previous checkpoint $\theta_{s-1}$ and minimizes $\mathcal{L}_s$, so the operator is updated rather than retrained.

An accuracy gate closes each stage. Minimizing $\mathcal{L}_s$ yields a candidate $\theta_s$, and the candidate is evaluated on the frozen test set $\mathcal{T}_{s'}$ of every earlier prior, data unseen in any training stage. With $\varepsilon(\theta; \mathcal{T})$ the relative $L_2$ error of checkpoint $\theta$ on test set $\mathcal{T}$, the per-sounding $\lVert \hat{\mathbf{d}} - \mathbf{d} \rVert_2 / \lVert \mathbf{d} \rVert_2$ of the reconstructed decay averaged over the set, the candidate is accepted only if no earlier prior has regressed by more than a tolerance $\delta$,
\begin{equation}
    \text{accept } \theta_s \iff \varepsilon\big(\theta_s;\ \mathcal{T}_{s'}\big) \le \max\big\{ (1+\delta)\,\varepsilon^{\min}_{s'},\ \varepsilon^{\min}_{s'} + \varepsilon_{\mathrm{abs}} \big\} \quad \text{for every } s' < s,
    \label{eq:cigate}
\end{equation}
where $\varepsilon^{\min}_{s'}$ is the lowest error any accepted checkpoint has reached on $\mathcal{T}_{s'}$, $\delta = 0.25$, and $\varepsilon_{\mathrm{abs}}$ is one percentage point, so that a regression is rejected only when it is both relatively and absolutely large; on the deployed baselines the relative branch binds. A rejected candidate rolls back to $\theta_{s-1}$ and is retried once with the memory $M$ and the distillation weight $\lambda_d$ both doubled. If the retry also regresses, the deployed surrogate is unchanged.

\subsection{Validity check at inference}
\label{subsec:validity}
The surrogate returns a decay for any input, including a conductivity field far from every training prior. A validity check warns of the extrapolation, per sounding, at inference time.

The check is built on a deep ensemble \citep{lakshminarayanan2017simple}, $M_e = 5$ copies of the surrogate trained from different random initializations on the same data. Inside the training support the data pull all members to the same decay. Outside the support each member extrapolates on its own and the predictions spread. The disagreement score measures that spread, the standard deviation across members averaged over the gates,
\begin{equation}
    u(\sigma) \;=\; \frac{1}{n_g} \sum_{t=1}^{n_g} \sqrt{\frac{1}{M_e} \sum_{m=1}^{M_e} \Big( \hat{y}^{(m)}_{t} - \bar{y}_{t} \Big)^{2}}, \qquad
    \bar{y}_{t} = \frac{1}{M_e} \sum_{m=1}^{M_e} \hat{y}^{(m)}_{t},
    \label{eq:disagreement}
\end{equation}
where $\hat{y}^{(m)}_{t}$ is the prediction of member $m$ at gate $t$. The field check takes the unbiased $1/(M_e-1)$ in place of $1/M_e$ and averages over the gates the signal-to-noise screen retains; each threshold is calibrated and applied under its own convention, so the two are not on a comparable scale. At survey scale, where a sounding is inverted by ranking a shared prior ensemble, its score is the median of $u$ over the draws the ranking accepts.

A threshold turns the score into a decision. The threshold $u_{95}$ is the 95th percentile of $u$ on validation data of the training priors. A sounding with $u \le u_{95}$ is inverted with the surrogate, and a sounding with $u > u_{95}$ is routed to the full solver of Section~\ref{subsec:governing}. 

\subsection{Surrogate accelerated Bayesian inversion}
\label{subsec:sampler}
\begin{figure}[htbp]\centering
\includegraphics[width=\linewidth]{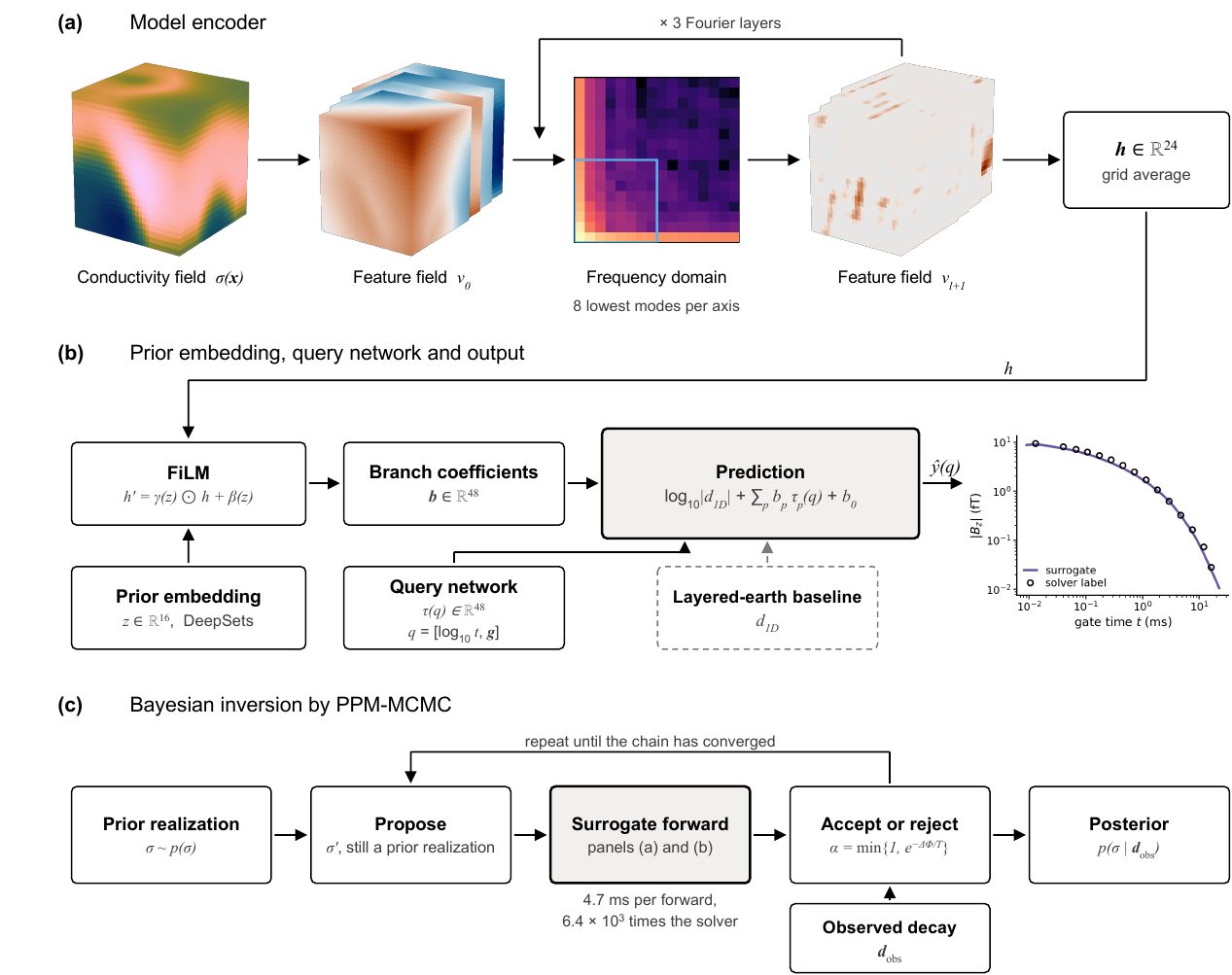}
    \caption{The neural operator surrogate and the sampler it drives. (a) The model encoder, with the eight lowest modes boxed. (b) The prior embedding, the query network and the output. The layered-earth baseline is dashed because the deployed checkpoints set it to zero. (c) The sampler. The gray fill marks the same operator, trained in (a) and (b) and called in (c).}
\label{fig:method}
\end{figure}

Bayesian inversion treats the conductivity field as a random quantity and recovers the posterior, $p(\sigma \mid \mathbf{d}_{\mathrm{obs}}) \propto p(\mathbf{d}_{\mathrm{obs}} \mid \sigma) \, p(\sigma)$, given the observed decay $\mathbf{d}_{\mathrm{obs}}$ of a sounding. The prior $p(\sigma)$ is the geostatistical prior, and the likelihood $p(\mathbf{d}_{\mathrm{obs}} \mid \sigma)$ scores how probable the observed decay is when the subsurface is $\sigma$, by simulating the decay of $\sigma$ and comparing it with the observation. The posterior has no closed form, so it is explored by Markov chain Monte Carlo (MCMC). The chain starts from a geological prior realization and repeats one step, propose a perturbed conductivity field, simulate its decay, and accept or reject the proposal by comparing its data fit with the current field, as laid out in Figure~\ref{fig:method} (c). After enough iterations, the accepted fields are distributed by the posterior \citep{metropolis1953equation, hastings1970monte, mosegaard1995monte}. One step needs three ingredients, a likelihood that scores the fit, a proposal that generates the perturbed field, and an acceptance rule. The following paragraphs define each in turn.

The likelihood is set by the measurement noise. Each observed data point is treated as the true decay plus a Gaussian error whose standard deviation is a relative term with an additive floor, the form in standard use for AEM \citep{green2003estimating},
\begin{equation}
    \varsigma_{t,n} = \rho \, \big| d^{\mathrm{obs}}_{t,n} \big| + \varsigma_0, \qquad \rho = 0.08, \quad \varsigma_0 = 0.01~\mathrm{fT},
    \label{eq:noise}
\end{equation}
where $d^{\mathrm{obs}}_{t,n}$ is the observed data at gate $t$ of sounding $n$, $\rho$ is the relative error, and $\varsigma_0$ is the additive floor. The same model applies to the controlled and the field inversions, and it defines the signal-to-noise screen that drops gates whose median amplitude falls below the noise.

Under this Gaussian noise, the negative log of the likelihood is, up to a constant, a sum of squared residuals scaled by $\varsigma_{t,n}$. The data misfit the chain evaluates is that sum with one additional dynamic-range weight,
\begin{equation}
    \Phi(\sigma) = \frac{1}{n_g} \sum_{t=1}^{n_g} \omega_t^2 \, \frac{1}{N_d} \sum_{n=1}^{N_d} \left( \frac{\hat{d}_{t,n} - d^{\mathrm{obs}}_{t,n}}{\varsigma_{t,n}} \right)^{2}, \qquad
    \omega_t = \big( \max_n d^{\mathrm{obs}}_{t,n} - \min_n d^{\mathrm{obs}}_{t,n} \big)^{-1},
    \label{eq:misfit}
\end{equation}
where $\hat{d}_{t,n}$ is the predicted data of Equation~\ref{eq:deeponet}, from the surrogate in the accelerated inversions and from the solver in the reference runs, and $N_d$ is the number of soundings inverted together. The weight $\omega_t$ is the reciprocal dynamic range of gate $t$ across those soundings, so it down-weights the gates whose amplitude varies most across the set. Every inversion reported here is single-sounding, $N_d = 1$, for which $\omega_t \equiv 1$ and $\Phi$ reduces to the gate-averaged $\chi^2$. A lower $\Phi(\sigma)$ means a more probable field under the likelihood.

The proposal generates the perturbed field by the PPM \citep{caers2003history, caers2006probability}. A prior realization is generated from underlying uniform variables, and the sampler perturbs the uniforms rather than the field itself. Each uniform $w_0$ is moved toward an independent draw $w_1$ with radius $r$,
\begin{equation}
    w_r = \mathrm{PPM}(w_0, r, w_1) =
    \begin{cases}
        w_0, & r w_0 < w_1 < 1 - r + r w_0,\\[2pt]
        w_1 / r, & w_1 \le r w_0,\\[2pt]
        (w_1 + r - 1)/r, & \text{otherwise},
    \end{cases}
    \label{eq:ppm}
\end{equation}
where $r$ controls the step size, a small $r$ leaving most uniforms unchanged. The perturbed $w_r$ stays marginally uniform for any $r$, so every proposal remains a prior realization, making the method prior-agnostic. The prior $p(\sigma)$ enters the chain through this construction, and the kernel is symmetric, with no proposal-ratio correction.

The acceptance rule decides between the proposed and the current field. A proposal is accepted with the tempered Metropolis probability $\alpha = \min\{1, \exp(-(\Phi_r - \Phi_i)/T)\}$, where $\Phi_r$ and $\Phi_i$ are the proposed and current misfits and $T$ is the sampling temperature. A better fit is always accepted, and a worse fit is accepted with a probability that falls with the misfit increase, which lets the chain explore the posterior rather than descend to a single best model. A rejected large step of radius $r_1$ triggers a smaller delayed-rejection step of radius $r_2$ with the detailed-balance correction of \citet{green2001delayed}. The radius $r_2$ is annealed over the iterations and the temperature is adapted by Robbins-Monro toward the acceptance rate $\alpha^{*} = 0.23$ \citep{robbins1951stochastic, roberts1997weak}. The sampler is forward agnostic. The identical algorithm runs with the surrogate or with the solver, which is the basis of the equivalence test of Section~\ref{subsec:equiv}.

\section{Results}
\label{sec:results}

\subsection{Surrogate accuracy and validity}
\label{subsec:accuracy}

\begin{figure}[htbp]\centering
\includegraphics[width=\linewidth]{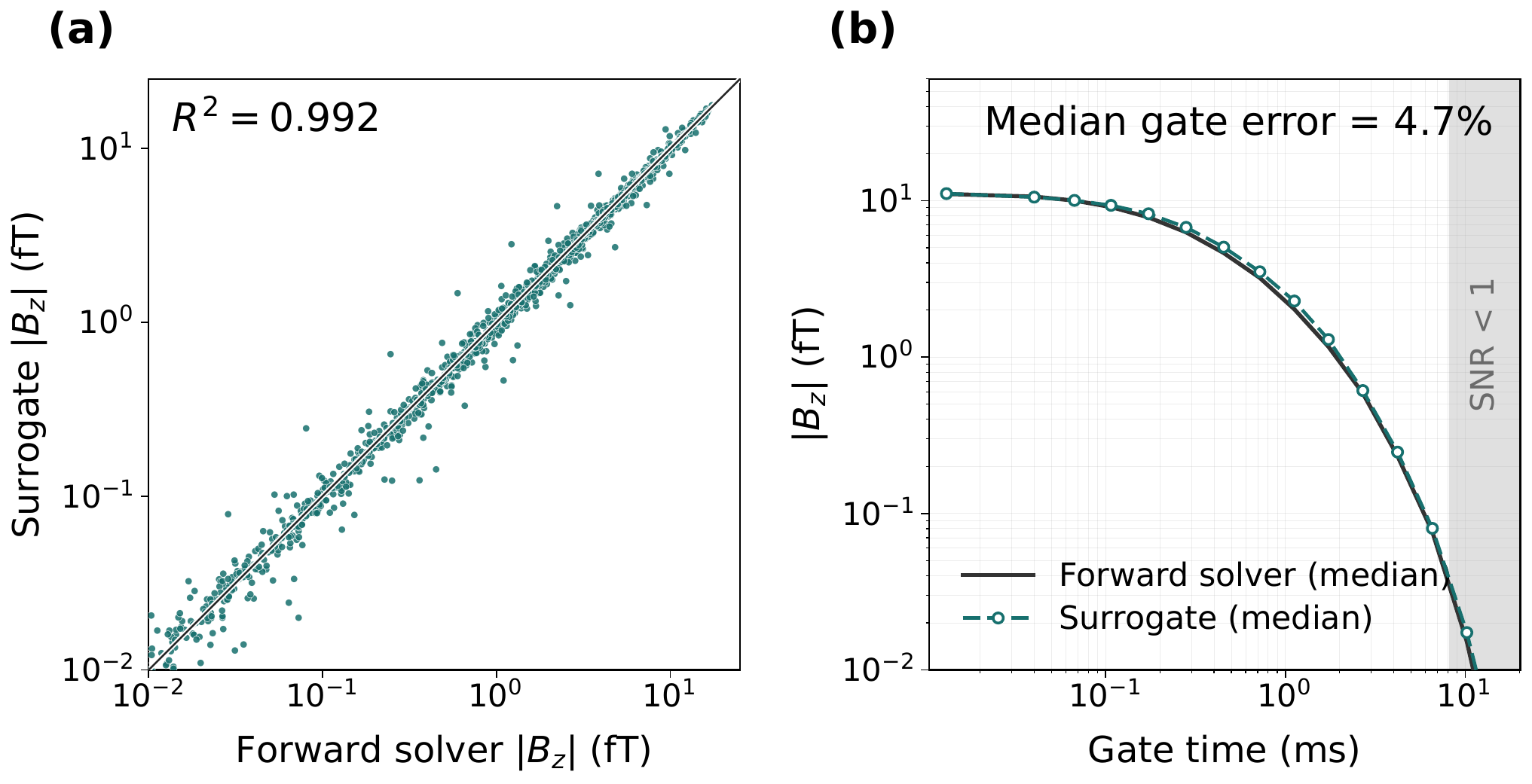}
    \caption{Forward accuracy on the Capricorn test data. (a) Surrogate against forward solver at the fifteen gates, with the 1:1 line. (b) The median sounding, surrogate dashed and forward solver solid. The shaded strip marks gates whose median decay falls below the noise model.}
\label{fig:accuracy}
\end{figure}

We evaluate the surrogate by comparison with the forward solver, since the surrogate aims to reproduce the solver up to its numerical accuracy. Figure~\ref{fig:accuracy}(a) compares the surrogate and the solver on the Capricorn test data, the 125 soundings not used in training, with $R^2=0.992$ on $\log_{10}|B_z|$ and a median gate error of 4.7\% for the initialization shown. Over ten initializations the error is $5.0 \pm 0.3$\%. Figure~\ref{fig:accuracy}(b) shows the median sounding, where the surrogate follows the solver across the full decay.

\begin{figure}[htbp]\centering
\includegraphics[width=0.62\linewidth]{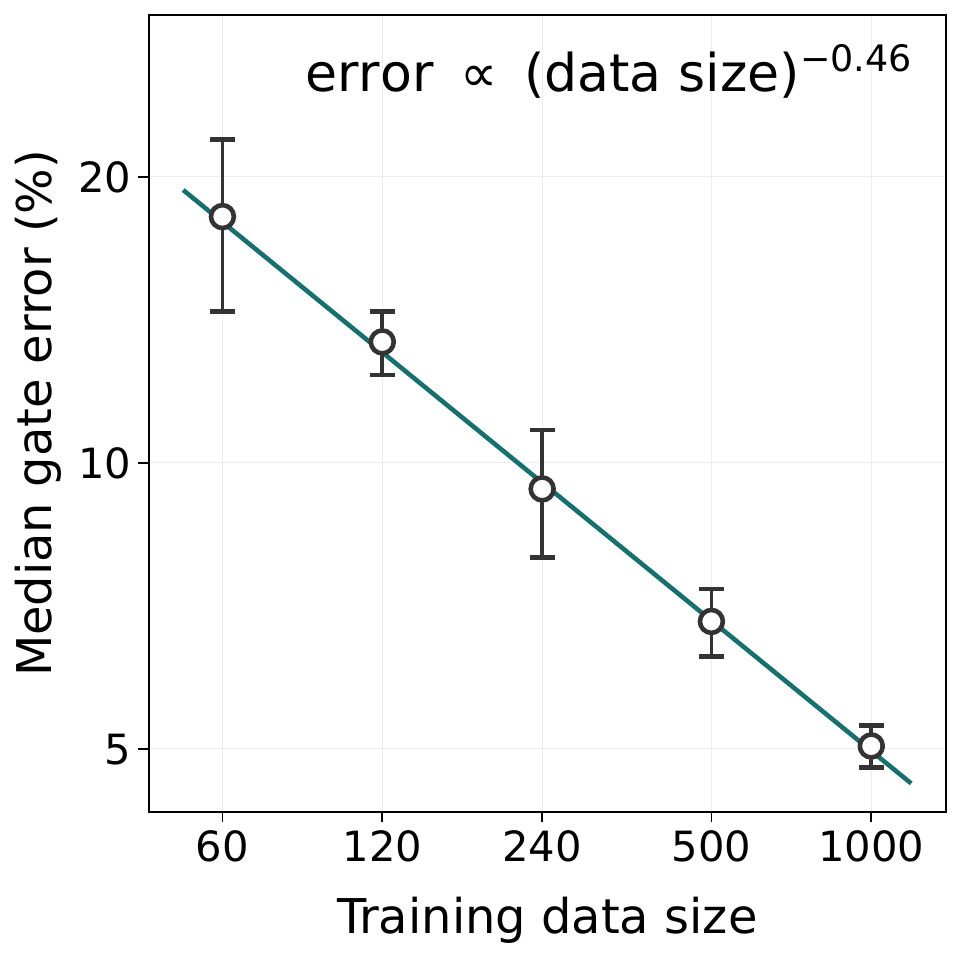}
    \caption{Test error against training set size on Capricorn. Markers and bars are the mean and standard deviation of ten runs at each size, scored on the same 125 test soundings. The line is a power law fitted to all fifty runs.}
\label{fig:lc}
\end{figure}

The accuracy is limited by the amount of training data rather than by the network capacity. Figure~\ref{fig:lc} shows the test error against the training data size on Capricorn, from 60 to 1{,}000 training soundings, ten runs at each size, scored on the same 125 test soundings. The error falls as a power law, $\varepsilon \propto n^{-0.46}$ in the number of training soundings $n$, and is still falling at the largest size we reach.

The surrogate error is small against the measurement noise the inversion already carries. Each gate is observed with the uncertainty of Equation~\ref{eq:noise}, and a surrogate error below that level keeps the misfit unchanged. In those units the error has a median of 0.28 standard deviations, and the last gate, whose relative error is largest, sits at 0.04 of a standard deviation because the 0.01\,fT floor supplies 98.8\% of its uncertainty.

\begin{figure}[htbp]\centering
\includegraphics[width=\linewidth]{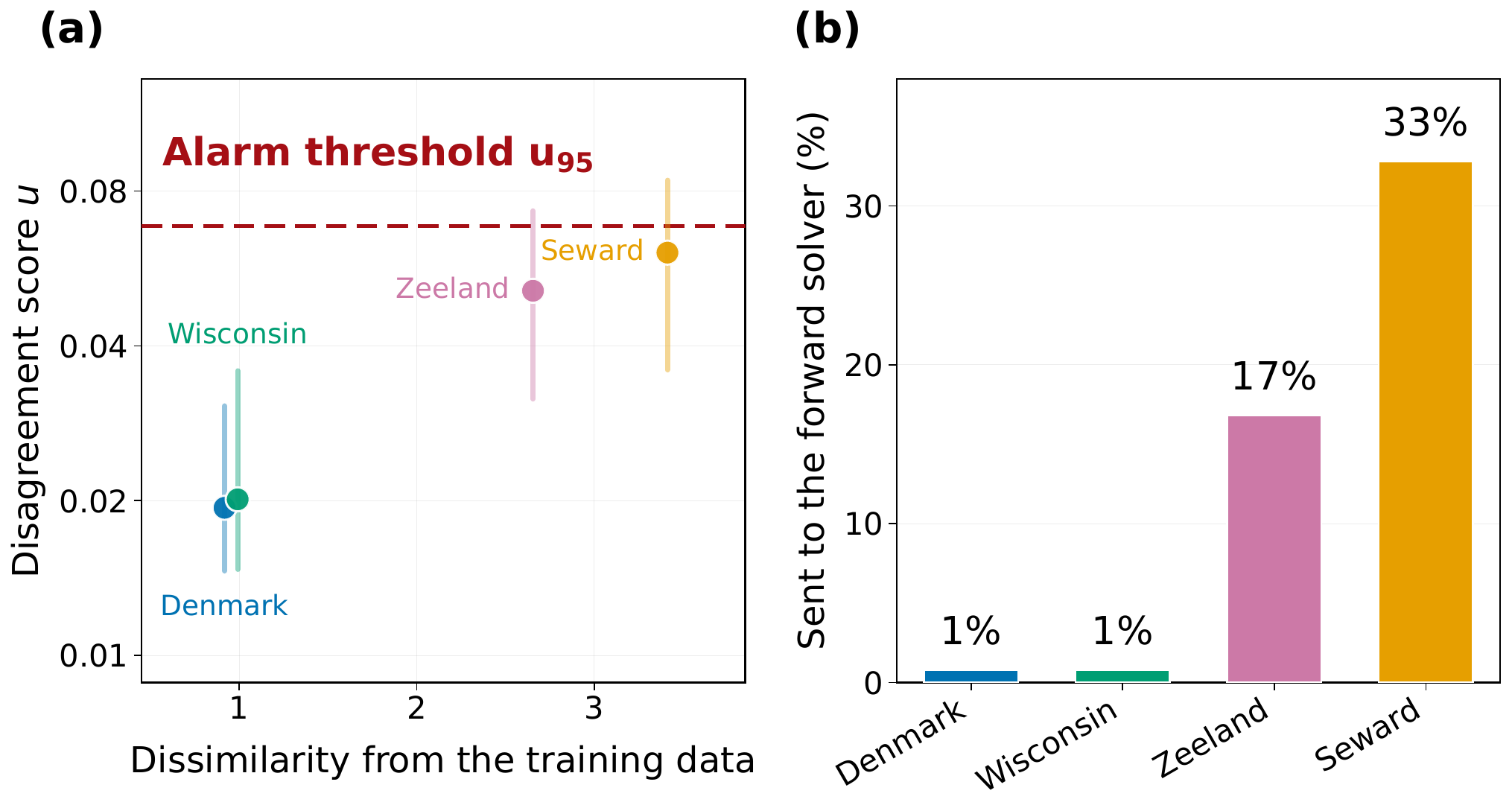}
    \caption{The validity check on the four regions the operator has not learned, trained on the Capricorn prior alone. (a) The disagreement score of Equation~\ref{eq:disagreement} against dissimilarity, the median distance from a test model to the nearest Capricorn training model in a standardized 64-number lateral profile. Markers are the median over soundings and bars run from the tenth to the ninetieth percentile. The dashed line is the threshold $u_{95}$, the 95th percentile of the score on the Capricorn validation data. (b) The fraction of each region's test soundings above the threshold, which are inverted with the forward solver.}
\label{fig:oodsyn}
\end{figure}

Once we have trained the surrogate on the Capricorn prior, exploration moves to geology that may be distinctly different, and nothing guarantees the accuracy carries over. We therefore score the trained operator, without further training, on the test data of the four other regions. Figure~\ref{fig:oodsyn}(a) orders the regions by how unlike the Capricorn training data they are. Denmark and Wisconsin are close in accuracy compared to Capricorn's own test soundings, 6.6\% and 5.4\% versus 5.2\% on Capricorn itself. Zeeland is well outside and the operator reaches 11.8\%. Seward is the exception at 67.7\%. The warning of this error comes from the disagreement score of Equation~\ref{eq:disagreement}, the spread among the five ensemble members that share the training data and differ only in initialization. Inside the training range the data pull the members to the same decay, and outside it each member extrapolates on its own. Figure~\ref{fig:oodsyn}(b) gives the fraction of each region the check sends to the solver, with the threshold at the 95th percentile of the score on the Capricorn validation data. The check warns on 32.8\% of Seward and 16.8\% of Zeeland, against 0.8\% of Denmark and Wisconsin and 4.0\% of the Capricorn test data, so it only flags those regions with poor prediction, and does not waste compute on cases that are already accurate. 

\subsection{Continual learning across priors}
\label{subsec:contres}

We train one surrogate on the five geological priors, sequentially, as would be the case when flying many surveys around the world. Here, we assume the Capricorn survey is flown first. The risk of a sequence is that the region already learned deteriorates. We therefore compare our surrogate with fine-tuning, both trained on the same data in the same order. Fine-tuning trains on each new region's training data alone. Our surrogate trains on the new region together with replay of stored models from the earlier priors and a distillation term that keeps the update close to the previous checkpoint's predictions. Every error below is the median gate error on the 125 test soundings of the region named, averaged over five initializations.

\begin{figure}[htbp]\centering
\includegraphics[width=\linewidth]{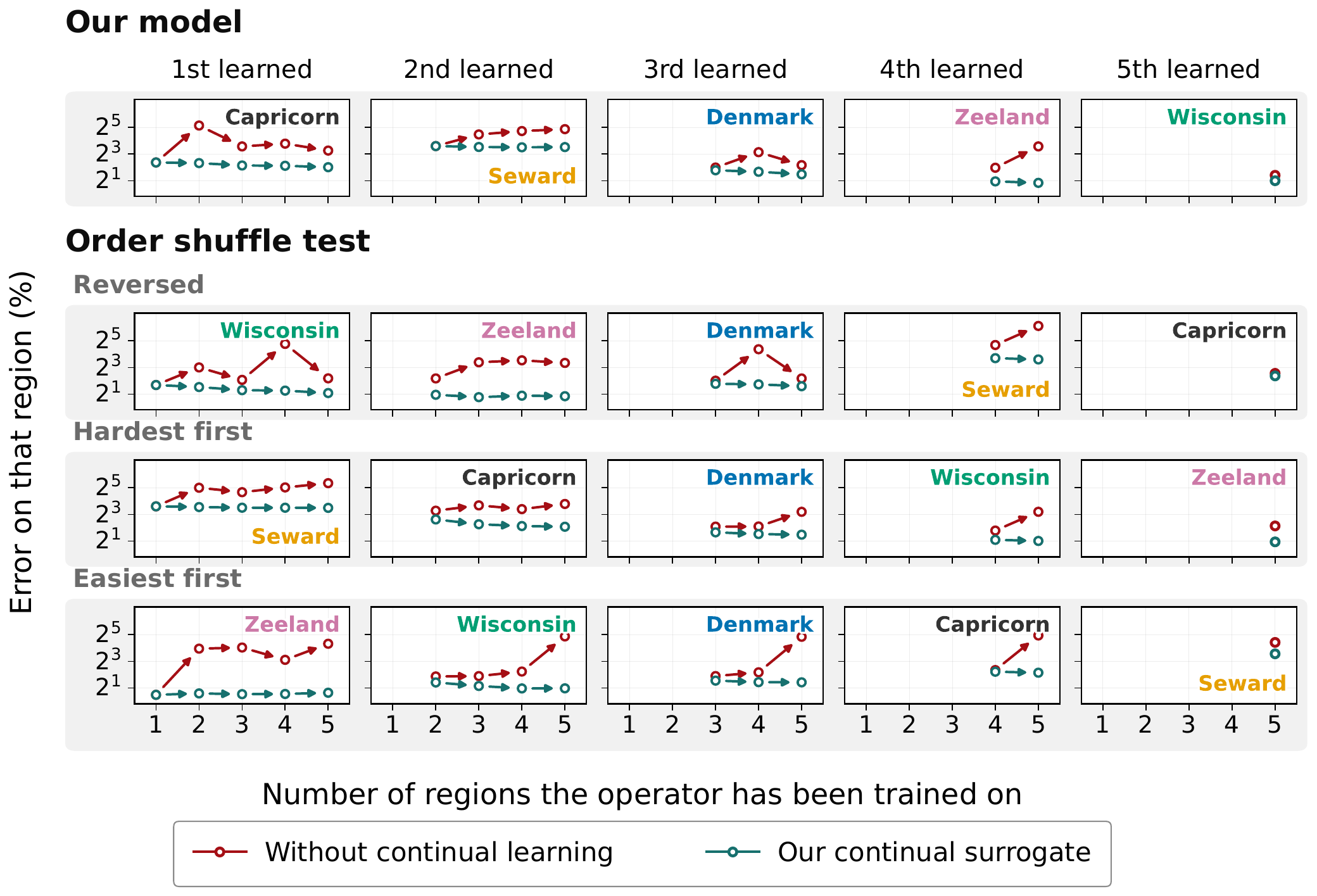}
    \caption{Continual learning with and without replay of the earlier priors, across the five geological priors and four arrival orders. Each row is one order, the order our surrogate was trained in first, and each column is a position in the sequence, with the region named in its panel. Arrows run from the stage that taught the region to the end of the sequence.}
\label{fig:forget}
\end{figure}

Figure~\ref{fig:forget} follows each region from the stage that taught it to the end of the sequence. Under fine-tuning the error on a region rises once the operator moves on, by 7.5 percentage points averaged over the four regions with a later stage. Capricorn goes from 5.2\% to 9.6\%, Seward from 12.1\% to 29.1\%, and Zeeland from 3.9\% to 11.9\%. Under our surrogate the same average is $-0.6$ percentage points, negative in every initialization. Each region ends the sequence more accurate than when the operator learned it. Capricorn has 5.2\% error as the only region the operator knows and 4.0\% once the other four are in, Seward 12.1\% against 11.5\%, and Denmark 3.5\% against 2.8\%. A new prior supplies conductivity fields the earlier ones did not contain, and the operator fits the same forward map over more of its domain.

Neither result depends on the order of the regions. We repeat the sequence in four arrival orders, the deployed one, its reverse, the hardest region first, and the easiest first, with three initializations for each new order, and Figure~\ref{fig:forget} gives all four. Our surrogate ends at 4.4, 4.9, 4.4 and 4.5\%, and its forgetting is negative in every order. Fine-tuning ends at 11.6, 18.9, 15.4 and 25.9\%, with forgetting from 7.5 to 23.6 percentage points. Retraining on all five regions at once reaches 4.6\%, against 4.4\% for the sequence.

\subsection{Equivalence of the surrogate and solver posteriors}
\label{subsec:equiv}

We test whether replacing the forward solver with the surrogate changes the posterior. The same Bayesian inversion runs twice on the same measured sounding, once with the solver computing every forward response and once with the surrogate, under the Capricorn prior and the operator of the continual sequence. The first comparison scores one pool of prior realizations twice. A pool of 400 realizations is forwarded at the recorded geometry, each realization is scored against the observed decay, and the best-fitting 10\% form the posterior, so only the forward operator differs between the two runs. Over fifteen soundings of line 1005801 the two posteriors share 85.5\% of the accepted realizations, at a data-fit residual of 11.3\% under the surrogate against 12.2\% under the solver, and per depth they differ by 0.45 of the distance two random halves of the same pool differ by. The surrogate moves the posterior less than the sampling does.

\begin{figure}[htbp]\centering
\includegraphics[width=0.62\linewidth]{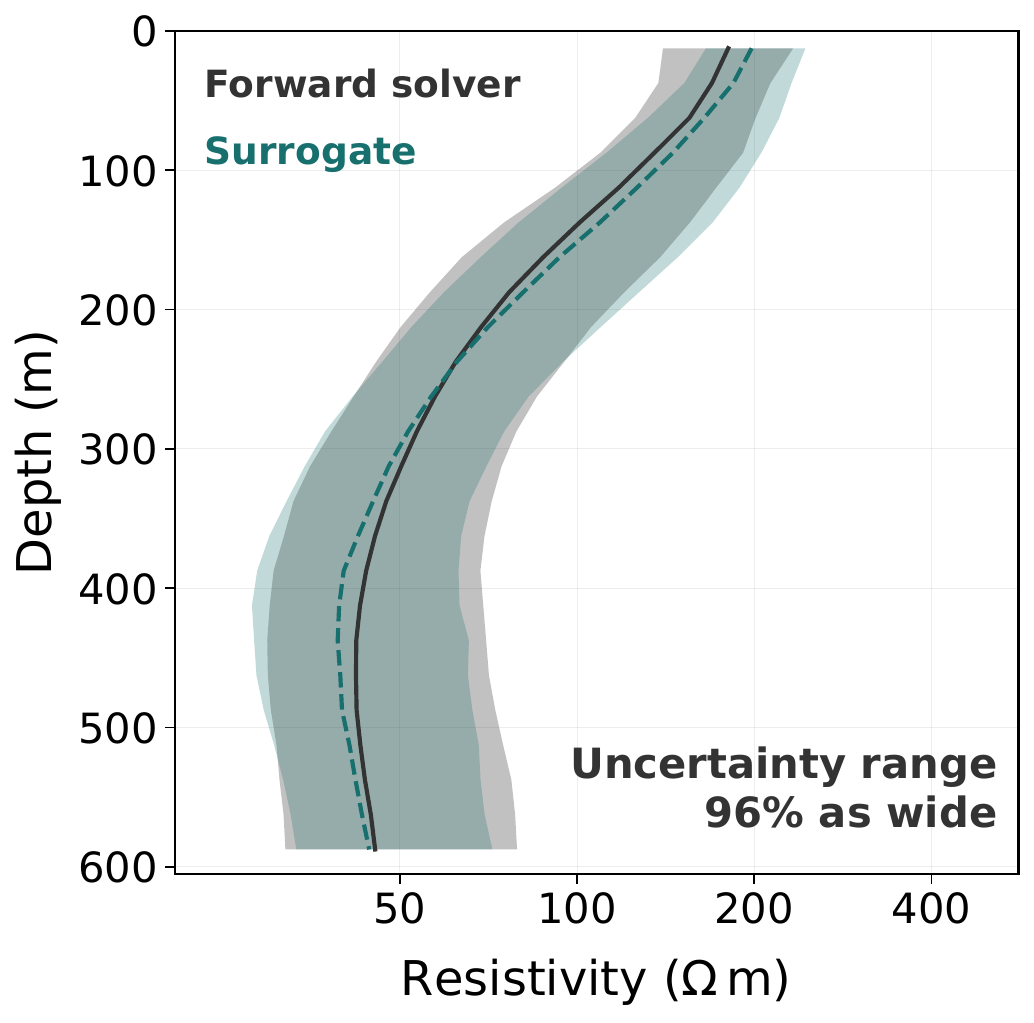}
    \caption{Sampler-level equivalence on the Capricorn prior. The identical PPM-MCMC sampler runs with the forward solver and with the surrogate, at the same budget and seeds, on one measured sounding of line 1005801. Bands are the 10--90\% posterior range of the volume-mean resistivity, lines the medians, pooled over thirty-two chains per operator. The quoted 96\% is the median over depth of the surrogate range divided by the solver range.}
\label{fig:mcmceq}
\end{figure}

The second comparison involves the full PPM-MCMC sampler. The identical sampler runs with each operator at the same iteration budget and the same random seeds, thirty-two chains of 500 iterations on one sounding of line 1005801, 8{,}000 retained states apiece. Figure~\ref{fig:mcmceq} shows the two posteriors. The uncertainty range the surrogate returns is 96\% as wide as the solver's, the two ranges share 85\% of their combined extent at the median depth, and the final acceptance is 0.30 against 0.28. The surrogate-driven and solver-driven posteriors differ by 10.3 times what a random split of the pooled states gives, and two independent halves of the solver run differ from each other by 12.2. Neither engine is fully mixed at this budget, with a Gelman-Rubin statistic of 1.39 for the solver and 1.47 for the surrogate, so the comparison bounds the surrogate's effect. The surrogate ran the experiment in 2{,}718\,s against 1{,}138{,}000\,s for the solver, forty-five minutes against thirteen days.

The surrogate posterior is calibrated, its uncertainty neither too narrow nor too wide. To test this, we take 400 realizations of the Capricorn prior the operator has not been trained on, treat each in turn as the truth, invert its stored solver response with the surrogate under the acceptance rule the survey inversion uses, and count how often the truth falls inside the posterior credible intervals. A 50\% interval contains the truth 50.6\% of the time, an 80\% interval 79.3\%, a 90\% interval 88.7\% and a 95\% interval 92.4\%, so the largest departure from nominal over the four levels is 2.6 percentage points and the smallest is 0.6. 

\subsection{Survey-scale inversion and prior updating}
\label{subsec:field}

We now scale from one sounding to the survey, the 2{,}155{,}272 soundings of the Capricorn TEMPEST data. Bayesian inversion at this scale is infeasible for the solver, whose forward evaluations alone would take tens of thousands of years. With the surrogate the inversion costs seconds, and testing the geological prior against the survey costs minutes. We first test the prior against every flown line, then invert the survey under it, and revise the prior where the test fails.

\begin{figure}[htbp]\centering
\includegraphics[width=\linewidth]{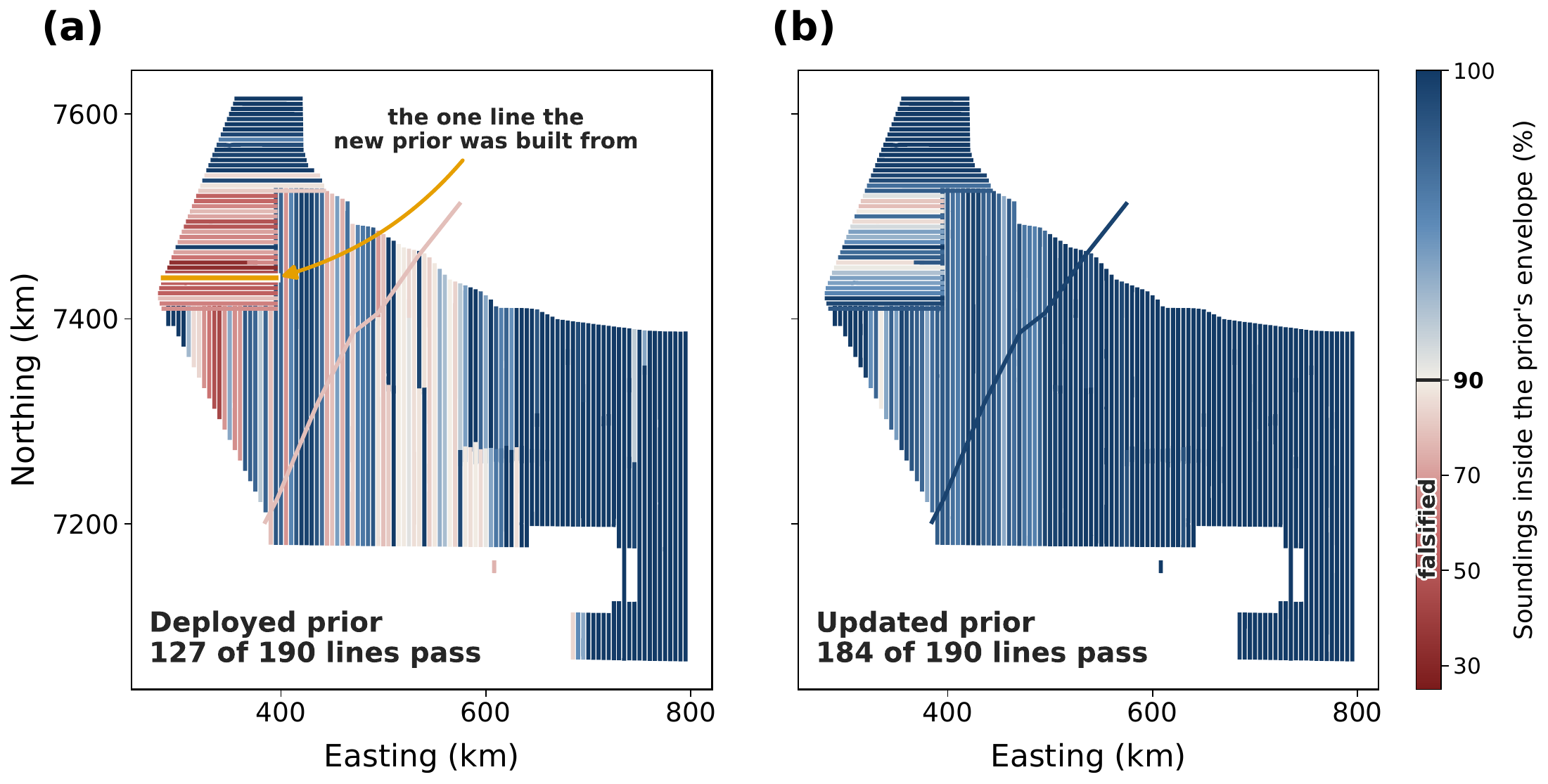}
    \caption{The falsification test over the whole survey, before and after one turn of the prior revision. Each flown line is colored by its coverage, the fraction of its soundings falling inside the 95\% envelope of the prior's forward responses, and a line is falsified below the 90\% level marked on the bar. (a) The deployed Capricorn prior, with the line the updated prior was built from marked in amber. (b) The updated prior.}
\label{fig:loop}
\end{figure}

The falsification test asks whether the observed soundings look like members of the prior. The prior's forward responses define a 95\% envelope in data space, the 95th percentile of their robust Mahalanobis distance in the principal-component subspace carrying 95\% of their variance, and a line is falsified when fewer than 90\% of its soundings fall inside it. The Capricorn prior contains 92\% of its grounding line 1005801 and 94\% of a neighboring line it never saw. Figure~\ref{fig:loop}(a) extends the test to the 190 unseen lines. The prior explains 127 of them at a median coverage of 97.9\%, so a prior grounded on one line holds over most of a survey 500\,km across, and the 63 lines it does not explain form a block in the west and north-west, named before any inversion is run.

The surrogate then inverts the survey under this prior. Each sounding scores a shared ensemble of 10{,}000 prior realizations forwarded at its own recorded geometry, and the 100 best-fitting realizations form its posterior. The 2{,}131{,}667 soundings whose transmitter clearance does not exceed the 180\,m ceiling of the trained range are inverted in 24.9\,s on one GPU, at 11.7\,$\mu$s per sounding, against about 26{,}300 years for the solver at 38.8\,s per forward solve. The validity check flags no sounding inside the trained 90 to 180\,m clearance band, so the surrogate inverts the entire survey without a fall-back to the solver. Resistivity rises with depth at 84\% of soundings, from a median of 63\,$\Omega\cdot$m at 38\,m to 340\,$\Omega\cdot$m at 238\,m, the conductive weathered regolith \citep{anandpaine2002} over the resistive banded iron formation \citep{dentith2014geophysics}, and the north-western block is the most resistive at depth of the whole survey, 929\,$\Omega\cdot$m against 264\,$\Omega\cdot$m elsewhere. 

\begin{figure}[htbp]\centering
\includegraphics[width=\linewidth]{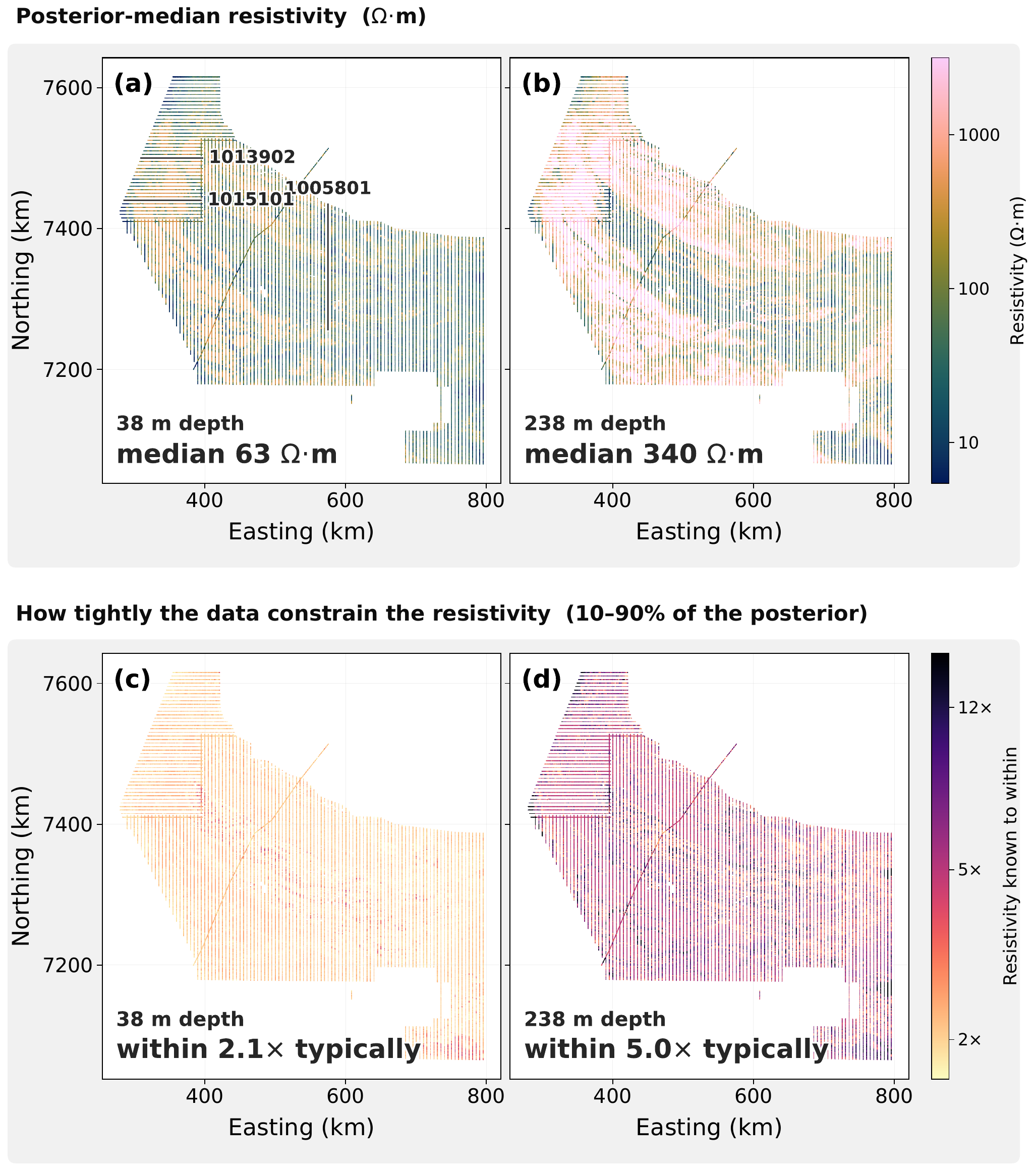}
    \caption{Whole-survey inversion of Capricorn TEMPEST. The top row is the posterior-median resistivity and the bottom row the width of the posterior, at 38\,m (a, c) and 238\,m (b, d). The width is the factor in resistivity between the 10th and 90th percentiles, so a value of 2 fixes the resistivity to within a factor of two. All 2{,}131{,}667 inverted soundings are shown, plotted one in four for legibility. Panel (a) marks line 1005801, which grounds the prior, and lines 1015101 and 1013902, which it does not explain.}
\label{fig:survey}
\end{figure}

Figure~\ref{fig:survey} maps the posterior-median resistivity and the width of the posterior. The resistivity is fixed to within a factor of 2.1 at 38\,m and 5.0 at 238\,m.

Line 1015101 shows the clearest failure, with the prior containing 59\% of its soundings, and we run one turn of the prior revision on it. Coverage alone would not settle the comparison, because widening any prior raises coverage, so the control is the deployed prior widened by the same amount without moving its center. The control grows the model space from 7.90 to 10.53 decades and fixes 31 of the 63 falsified lines while losing 19 the deployed prior already explained. A two-component mixture that correlates the layers, on the same layer ranges, fixes 57 and loses none, and the survey increases from 127 to 184 of 190 lines in 8.65 decades (Table~\ref{tab:variants}, Figure~\ref{fig:loop}(b)). The gain comes from the layer correlation and not from width. Folding the new prior into the operator repeats the continual step on measured data. Its error on the new prior falls from 28.1\% to 8.1\%, the accuracy on the earlier region is unchanged, and the posterior on the original line does not move. One flight line and one retraining revise the prior for a survey 500\,km across.

\begin{table}[htbp]\centering
    \caption{The four priors compared. Model space is the width of the prior in decades of $\log_{10}\sigma$ summed over layers, for the mixture the weighted mean of its two components. Lines explained and lines lost are counted over the 190 unseen flight lines.}
\label{tab:variants}
\footnotesize
\begin{tabular}{lccc}
\toprule
prior & model space (decades) & lines explained (of 190) & lines lost \\
\midrule
deployed Capricorn           & 7.90  & 127 & --  \\
width control                & 10.53 & 139 & 19 \\
regional, independent layers & 11.75 & 176 & 4 \\
two-component mixture        & 8.65  & 184 & 0 \\
\bottomrule
\end{tabular}
\end{table}

\section{Discussion}
\label{sec:discussion}

The continual learning result follows from the invariance of the operator. The physics does not change with the prior, so a new prior shifts the inputs the surrogate sees. Replay with distillation already contains every region so far learned. In fact, we find the accuracy on earlier priors improves as continual learning progresses. Across four arrival orders the operator ends within half a percentage point of itself, where fine-tuning spreads over fifteen. This closes the distribution-shift limitation left open by static surrogate frameworks \citep{liu2026probability}.

The framework changes the cost structure of probabilistic inversion. A Bayesian product over two million soundings, infeasible for the solver at any budget, costs seconds, and testing a prior against an entire survey costs minutes. The prior stops being a fixed input chosen before the inversion and becomes a quantity the survey itself revises, and the falsification test tells where the next prior is needed at the cost of one training data set and one retraining. 

Two diagnostics carry beyond the Capricorn application. The disagreement score is a detector of leaving the training support. As ensemble members become individually more accurate they become more similar, and the partial correlation of score with error at fixed clearance falls from $+0.485$ under an earlier, less accurate ensemble to $-0.014$, so the score degrades as the ensemble improves. The second diagnostic is a control for prior updating. Coverage rises whenever a prior is widened, so a coverage gain is evidence of a better prior only against the same prior widened by the same amount without re-centering. Here the widened control occupies the largest model space of the four priors tried and still explains fewer lines than the mixture, and it falsifies nineteen lines the deployed prior had explained.

\section{Conclusion}
\label{sec:conclusion}

We have developed a continually learning neural-operator surrogate of the three-dimensional airborne electromagnetic forward operator, with a validity check that scores each sounding against the support the operator was trained on.

The operator learns a sequence of geological priors without forgetting, and replay of stored models from the earlier priors does more than that. Every prior ends the sequence more accurate than when the operator learned it, because a new prior provides conductivity models the earlier ones did not contain and the operator fits the same forward map over more of its domain. That gain survives every arrival order we tested.

The surrogate reproduces the solver posterior. Driven by the identical sampler at matched settings, it departs from the solver by less than two independent solver-driven chains depart from each other, and its credible intervals cover the truth within 2.6 percentage points of nominal. The validity check routes out-of-range soundings to the solver.

The prior becomes a testable and revisable input rather than a fixed choice. A cheap forward ensemble tests the prior against every line of a survey before any inversion is run. A second prior built from a single failing line, folded into the same operator, raises the lines the survey explains from 127 to 184 of 190 and falsifies none the first prior already explained, so what the data ask for is the structure of the prior and not its width.

After the one-time cost of the training data, a sounding costs 11.7\,$\mu$s, and inverting the survey's two million soundings takes seconds against about 26{,}300 years for the solver. Probabilistic inversion, with calibrated uncertainty, is available at the scale of an airborne survey. The framework is the foundation for a joint three-dimensional magnetics and electromagnetic inversion developed separately.

\section*{Data and code availability}
The Capricorn TEMPEST survey data are publicly available from Geoscience Australia at \url{https://pid.geoscience.gov.au/dataset/ga/81642} (eCat record 81642), including the acquisition and processing report \citep{cgg2014capricorn}. The sources of the geological priors are cited in Table~\ref{tab:regions}. The code will be public on Github upon acceptance for publication.

\bibliographystyle{plainnat}
\bibliography{references}

@article{anandpaine2002,
  author={Anand, Ravi R and Paine, Marko},
  year    = {2002},
  title   = {Regolith geology of the {Yilgarn Craton}, {Western Australia}: implications for exploration},
  journal = {Australian Journal of Earth Sciences},
  volume  = {49},
  number  = {1},
  pages   = {3--162}
}

@article{auken2015overview,
  title={An overview of a highly versatile forward and stable inverse algorithm for airborne, ground-based and borehole electromagnetic and electric data},
  author={Auken, Esben and Christiansen, Anders Vest and Kirkegaard, Casper and Fiandaca, Gianluca and Schamper, Cyril and Behroozmand, Ahmad Ali and Binley, Andrew and Nielsen, Emil and Effers{\o}, Flemming and Christensen, Niels B{\o}ie and others},
  journal={Exploration Geophysics},
  volume={46},
  number={3},
  pages={223--235},
  year={2015},
  publisher={Australian Society of Exploration Geophysicists}
}

@article{barfod2016compiling,
  title={Compiling a national resistivity atlas of {Denmark} based on airborne and ground-based transient electromagnetic data},
  author={Barfod, Adrian AS and M{\o}ller, Ingelise and Christiansen, Anders V},
  journal={Journal of Applied Geophysics},
  volume={134},
  pages={199--209},
  year={2016},
  publisher={Elsevier}
}

@article{caers2003history,
  title={History matching under training-image-based geological model constraints},
  author={Caers, Jef},
  journal={SPE Journal},
  volume={8},
  number={03},
  pages={218--226},
  year={2003},
  publisher={OnePetro}
}

@article{caers2006probability,
  title={The Probability Perturbation Method: A New Look at {Bayesian} Inverse Modeling: {Caers} and {Hoffman}},
  author={Caers, Jef and Hoffman, Todd},
  journal={Mathematical Geology},
  volume={38},
  number={1},
  pages={81--100},
  year={2006},
  publisher={Springer}
}

@article{caers2026stochastic,
  title={Stochastic Inversion of geophysical data by sequential {Bayesian} updating under a non-stationary {Gaussian} process prior},
  author={Caers, Jef and Li, Peng and Kloeckner, Jonas and Daza, Juan Pablo and Yin, Zhen and Scheidt, Celine},
  year={2026},
  journal={EarthArXiv preprint},
  publisher={EarthArXiv}
}

@article{cockett2015simpeg,
  title={{SimPEG}: An open source framework for simulation and gradient based parameter estimation in geophysical applications},
  author={Cockett, Rowan and Kang, Seogi and Heagy, Lindsey J and Pidlisecky, Adam and Oldenburg, Douglas W},
  journal={Computers \& Geosciences},
  volume={85},
  pages={142--154},
  year={2015},
  publisher={Elsevier}
}

@book{dentith2014geophysics,
  author    = {Dentith, Michael and Mudge, Stephen T.},
  title     = {Geophysics for the Mineral Exploration Geoscientist},
  publisher = {Cambridge University Press},
  year      = {2014},
  address   = {Cambridge},
  isbn      = {9780521809511},
  doi       = {10.1017/CBO9781139024358}
}

@techreport{Emond2024Seward,
  author      = {Emond, A. M. and Fusso, L. A. and {SkyTEM Canada Incorporated}},
  title       = {{Seward Peninsula} airborne electromagnetic survey, {Kigluaik}, {Bendeleben}, and {Darby} mountains, {Alaska}},
  institution = {Alaska Division of Geological \& Geophysical Surveys},
  type        = {Geophysical Report},
  number      = {GPR 2024-2},
  year        = {2024},
  pages       = {9},
  doi         = {10.14509/31303},
  url         = {https://doi.org/10.14509/31303}
}

@article{roberts1997weak,
  title={Weak convergence and optimal scaling of random walk {Metropolis} algorithms},
  author={Roberts, Gareth O and Gelman, Andrew and Gilks, Walter R},
  journal={The Annals of Applied Probability},
  volume={7},
  number={1},
  pages={110--120},
  year={1997},
  publisher={Institute of Mathematical Statistics}
}

@article{green2003estimating,
  title={Estimating noise levels in {AEM} data},
  author={Green, Andy and Lane, Richard},
  journal={ASEG Extended Abstracts},
  volume={2003},
  number={2},
  pages={1--5},
  year={2003},
  publisher={Taylor \& Francis}
}

@article{green2001delayed,
  title={Delayed rejection in reversible jump {Metropolis--Hastings}},
  author={Green, Peter J and Mira, Antonietta},
  journal={Biometrika},
  volume={88},
  number={4},
  pages={1035--1053},
  year={2001},
  publisher={Biometrika Trust}
}

@article{hastings1970monte,
  title={{Monte Carlo} sampling methods using {Markov} chains and their applications},
  author={Hastings, W Keith},
  journal={Biometrika},
  volume={57},
  number={1},
  pages={97--109},
  year={1970},
  publisher={Oxford University Press}
}

@article{hinton2015distilling,
  title={Distilling the knowledge in a neural network},
  author={Hinton, Geoffrey and Vinyals, Oriol and Dean, Jeff},
  journal={arXiv preprint arXiv:1503.02531},
  year={2015}
}

@article{kirkpatrick2017overcoming,
  title={Overcoming catastrophic forgetting in neural networks},
  author={Kirkpatrick, James and Pascanu, Razvan and Rabinowitz, Neil and Veness, Joel and Desjardins, Guillaume and Rusu, Andrei A and Milan, Kieran and Quan, John and Ramalho, Tiago and Grabska-Barwinska, Agnieszka and others},
  journal={Proceedings of the National Academy of Sciences},
  volume={114},
  number={13},
  pages={3521--3526},
  year={2017},
  publisher={National Academy of Sciences}
}

@article{kitanidis1996geostatistical,
  title={On the geostatistical approach to the inverse problem},
  author={Kitanidis, Peter K},
  journal={Advances in Water Resources},
  volume={19},
  number={6},
  pages={333--342},
  year={1996},
  publisher={Elsevier}
}

@article{kovachki2023neural,
  title={Neural operator: Learning maps between function spaces with applications to {PDE}s},
  author={Kovachki, Nikola and Li, Zongyi and Liu, Burigede and Azizzadenesheli, Kamyar and Bhattacharya, Kaushik and Stuart, Andrew and Anandkumar, Anima},
  journal={Journal of Machine Learning Research},
  volume={24},
  number={89},
  pages={1--97},
  year={2023}
}

@article{krapez1996sequence,
  title={Sequence stratigraphic concepts applied to the identification of basin-filling rhythms in {Precambrian} successions},
  author={Krapez, B},
  journal={Australian Journal of Earth Sciences},
  volume={43},
  number={4},
  pages={355--380},
  year={1996},
  publisher={Taylor \& Francis}
}

@article{lakshminarayanan2017simple,
  title={Simple and scalable predictive uncertainty estimation using deep ensembles},
  author={Lakshminarayanan, Balaji and Pritzel, Alexander and Blundell, Charles},
  journal={Advances in Neural Information Processing Systems},
  volume={30},
  year={2017}
}

@article{li2017learning,
  title={Learning without forgetting},
  author={Li, Zhizhong and Hoiem, Derek},
  journal={IEEE Transactions on Pattern Analysis and Machine Intelligence},
  volume={40},
  number={12},
  pages={2935--2947},
  year={2017},
  publisher={IEEE}
}

@article{li2020fourier,
  title={Fourier neural operator for parametric partial differential equations},
  author={Li, Zongyi and Kovachki, Nikola and Azizzadenesheli, Kamyar and Liu, Burigede and Bhattacharya, Kaushik and Stuart, Andrew and Anandkumar, Anima},
  journal={arXiv preprint arXiv:2010.08895},
  year={2020}
}

@article{liu2026probability,
  title={Probability-Perturbation {McMC} for {3D} {Bayesian TDEM} Inversion with Unknown Covariance Parameters},
  author={Liu, Zhuo and Yin, Zhen and Kloeckner, Jonas and Muir, Jack and Caers, Jef},
  year={2026},
  journal={EarthArXiv preprint},
  publisher={EarthArXiv}
}

@article{lopez2017gradient,
  title={Gradient episodic memory for continual learning},
  author={Lopez-Paz, David and Ranzato, Marc'Aurelio},
  journal={Advances in Neural Information Processing Systems},
  volume={30},
  year={2017}
}

@article{lu2021learning,
  title={Learning nonlinear operators via {DeepONet} based on the universal approximation theorem of operators},
  author={Lu, Lu and Jin, Pengzhan and Pang, Guofei and Zhang, Zhongqiang and Karniadakis, George Em},
  journal={Nature Machine Intelligence},
  volume={3},
  number={3},
  pages={218--229},
  year={2021},
  publisher={Nature Publishing Group UK London}
}

@article{metropolis1953equation,
  title={Equation of state calculations by fast computing machines},
  author={Metropolis, Nicholas and Rosenbluth, Arianna W and Rosenbluth, Marshall N and Teller, Augusta H and Teller, Edward},
  journal={The Journal of Chemical Physics},
  volume={21},
  number={6},
  pages={1087--1092},
  year={1953},
  publisher={American Institute of Physics}
}

@article{minsley2022airborne,
  title={Airborne electromagnetic and magnetic survey data, northeast {Wisconsin} (ver. 1.1, {June} 2022)},
  author={Minsley, Burke J and Bloss, Ben B and Hart, David J and Fitzpatrick, William and Muldoon, Maureen A and Stewart, Esther K and Hunt, Randall J and James, Stephanie R and Foks, N Leon and Komiskey, Matthew J},
  journal={US Geological Survey (USGS) Data Release},
  pages={37},
  year={2022}
}

@article{moghadas2020soil,
  title={Soil electrical conductivity imaging using a neural network-based forward solver: Applied to large-scale {Bayesian} electromagnetic inversion},
  author={Moghadas, Davood and Behroozmand, Ahmad A and Christiansen, Anders Vest},
  journal={Journal of Applied Geophysics},
  volume={176},
  pages={104012},
  year={2020},
  publisher={Elsevier}
}

@article{moller2009integrated,
  title={Integrated management and utilization of hydrogeophysical data on a national scale},
  author={M{\o}ller, Ingelise and S{\o}ndergaard, Verner H and J{\o}rgensen, Flemming and Auken, Esben and Christiansen, Anders V},
  journal={Near Surface Geophysics},
  volume={7},
  number={5-6},
  pages={647--659},
  year={2009},
  publisher={Wiley Online Library}
}

@article{mosegaard1995monte,
  title={{Monte Carlo} sampling of solutions to inverse problems},
  author={Mosegaard, Klaus and Tarantola, Albert},
  journal={Journal of Geophysical Research: Solid Earth},
  volume={100},
  number={B7},
  pages={12431--12447},
  year={1995},
  publisher={Wiley Online Library}
}

@article{nabighian1979quasi,
  title={Quasi-static transient response of a conducting half-space—{An} approximate representation},
  author={Nabighian, Misac N},
  journal={Geophysics},
  volume={44},
  number={10},
  pages={1700--1705},
  year={1979},
  publisher={Society of Exploration Geophysicists}
}

@article{revil2017complex,
  title={Complex conductivity of soils},
  author={Revil, Andr{\'e} and Coperey, A and Shao, Z and Florsch, N and Fabricius, Ida Lykke and Deng, Y and Delsman, JR and Pauw, PS and Karaoulis, M and De Louw, PGB and others},
  journal={Water Resources Research},
  volume={53},
  number={8},
  pages={7121--7147},
  year={2017},
  publisher={Wiley Online Library}
}

@article{robbins1951stochastic,
  title={A stochastic approximation method},
  author={Robbins, Herbert and Monro, Sutton},
  journal={The Annals of Mathematical Statistics},
  pages={400--407},
  year={1951},
  publisher={JSTOR}
}

@article{sambridge2002monte,
  title={Monte {Carlo} methods in geophysical inverse problems},
  author={Sambridge, Malcolm and Mosegaard, Klaus},
  journal={Reviews of Geophysics},
  volume={40},
  number={3},
  pages={3--1},
  year={2002},
  publisher={Wiley Online Library}
}

@book{tarantola2005inverse,
  title={Inverse Problem Theory and Methods for Model Parameter Estimation},
  author={Tarantola, Albert},
  year={2005},
  publisher={SIAM}
}

@article{tarantola1982generalized,
  title={Generalized nonlinear inverse problems solved using the least squares criterion},
  author={Tarantola, Albert and Valette, Bernard},
  journal={Reviews of Geophysics},
  volume={20},
  number={2},
  pages={219--232},
  year={1982},
  publisher={Wiley Online Library}
}

@book{thorne1996geology,
  title={Geology of the {Rocklea} 1: 100,000 sheet},
  author={Thorne, AM and Tyler, IM},
  year={1996},
  publisher={Geological Survey of Western Australia}
}

@article{wang2022hierarchical,
  title={Hierarchical {Bayesian} inversion of global variables and large-scale spatial fields},
  author={Wang, Lijing and Kitanidis, Peter K and Caers, Jef},
  journal={Water Resources Research},
  volume={58},
  number={5},
  pages={e2021WR031610},
  year={2022},
  publisher={Wiley Online Library}
}

@article{wu2023fast,
  title={Fast {Bayesian} inversion of airborne electromagnetic data based on the invertible neural network},
  author={Wu, Sihong and Huang, Qinghua and Zhao, Li},
  journal={IEEE Transactions on Geoscience and Remote Sensing},
  volume={61},
  pages={1--11},
  year={2023},
  publisher={IEEE}
}

@article{yang2012three,
  title={Three-dimensional inversion of airborne time-domain electromagnetic data with applications to a porphyry deposit},
  author={Yang, Dikun and Oldenburg, Douglas W},
  journal={Geophysics},
  volume={77},
  number={2},
  pages={B23--B34},
  year={2012},
  publisher={Society of Exploration Geophysicists}
}

@article{yang2024efficient,
  title={An efficient {Bayesian} inference for geo-electromagnetic data inversion based on surrogate modeling with adaptive sampling {DNN}},
  author={Yang, Xu and Liu, Yunhe and Su, Yang and Yin, Changchun and Wang, Luyuan and Gao, Yu and Ren, Xiuyan and Zhang, Bo},
  journal={IEEE Transactions on Geoscience and Remote Sensing},
  volume={62},
  pages={1--17},
  year={2024},
  publisher={IEEE}
}

@article{rolnick2019experience,
  title={Experience replay for continual learning},
  author={Rolnick, David and Ahuja, Arun and Schwarz, Jonathan and Lillicrap, Timothy and Wayne, Gregory},
  journal={Advances in neural information processing systems},
  volume={32},
  year={2019}
}

@misc{cgg2014capricorn,
  author       = {{Geoscience Australia}},
  title        = {{AusAEM 02 WA/NT, 2019-20 Airborne Electromagnetic Survey: TEMPEST(R) airborne electromagnetic data and conductivity estimates}},
  year         = {2020},
  howpublished = {Geoscience Australia Data Release},
  publisher    = {Commonwealth of Australia (Geoscience Australia)},
  address      = {Canberra, Australia},
  doi          = {10.26186/144498},
  url          = {https://ecat.ga.gov.au/geonetwork/srv/api/records/8e598964-b4f3-4500-86ba-4c36d762f14e},
}

\end{document}